\RequirePackage{ifpdf}
\documentclass[12pt,letterpaper]{JHEP3}         
\usepackage{amssymb,amsfonts}

\usepackage[vcentermath]{youngtab}
\usepackage{wrapfig}
\usepackage{epsfig}

\def\be{\begin{eqnarray}}
\def\ee{\end{eqnarray}}
\newcommand{\nn}{\nonumber}
\newcommand\para{\paragraph{}}

\newcommand{\eqn}[1]{(\ref{#1})}

\newcommand{\ket}[1]{|{#1}\rangle}
\newcommand{\bra}[1]{\langle{#1}|}

\newcommand{\up}{\ket{\uparrow}}
\newcommand{\down}{\ket{\downarrow}}

\def\Dslash{\,\,{\raise.15ex\hbox{/}\mkern-12mu D}}
\def\Dbarslash{\,\,{\raise.15ex\hbox{/}\mkern-12mu {\bar D}}}
\def\delslash{\,\,{\raise.15ex\hbox{/}\mkern-9mu \partial}}
\def\delbarslash{\,\,{\raise.15ex\hbox{/}\mkern-9mu {\bar\partial}}}
\def\pslash{\,\,{\raise.15ex\hbox{/}\mkern-9mu p}}
\def\calDslash{\,\,{\raise.15ex\hbox{/}\mkern-12mu {\cal D}}}

\newcommand{\vp}{\varphi}
\newcommand{\tvp}{\tilde{\varphi}}
\newcommand{\tz}{\tilde{Z}}

\newcommand\tr{{\rm tr}}

\def\implies{\Rightarrow}

\def\lae{\mathrel{\mathop{\smash{\lower .5 ex \hbox{$\stackrel<\sim$}}}}}
\def\lae{\mathrel{\mathop{\smash{\lower .5 ex \hbox{$\stackrel>\sim$}}}}}

\title{ADHM and the 4d Quantum Hall Effect}

\author{Alec Barns-Graham, Nick Dorey, Nakarin Lohitsiri, David Tong and Carl Turner
\\
Department of Applied Mathematics and Theoretical Physics, \\
University of Cambridge, Cambridge, CB3 OWA, UK 
}

\abstract{Yang-Mills instantons are solitonic particles in $d=4+1$
  dimensional gauge theories. We construct and analyse the quantum
  Hall states that arise when these particles are 
restricted to the lowest Landau level. We describe the ground 
state wavefunctions for both Abelian and non-Abelian quantum Hall
states. Although our model is purely bosonic, we 
show that the excitations of this 4d quantum Hall state are
governed by the Nekrasov partition function of a certain five
dimensional supersymmetric gauge theory with Chern-Simons term. 
The partition function can also be interpreted as a   
variant of the Hilbert series of the instanton moduli space, counting 
holomorphic sections rather than holomorphic functions.  

\para
It is known that the Hilbert series of the instanton moduli space 
can be rewritten using mirror 
symmetry of 3d gauge theories in terms of Coulomb branch variables. 
We generalise this approach to include the effect of a five
dimensional Chern-Simons term. 
We demonstrate that the resulting Coulomb branch formula 
coincides with the corresponding Higgs branch Molien integral which, in
turn, reproduces the standard formula for the Nekrasov partition function.}

\begin{document}
\pagestyle{plain} \setcounter{page}{1}
\newcounter{bean}
\baselineskip16pt \setcounter{section}{0}

\newpage
\section{Introduction}

The purpose of this paper is to build a four-dimensional quantum Hall
state out of  Yang-Mills instantons. 

\para
This work has its motivation in condensed matter physics, but rapidly
converges towards a set of ideas and techniques that  are more familiar in
the world of  supersymmetry,  most notably the Nekrasov partition
function. Because of
these somewhat diverse ingredients, we will take this opportunity to
present some pertinent background material while summarising  our results.

\subsubsection*{Relationship to the 4d Quantum Hall Effect}

The two dimensional quantum Hall effect (QHE) is one of the gems of
physics, where beloved theoretical  ideas such as Chern-Simons
theories,  conformal symmetry, and the intricate structures that they contain, find
physical realisation in the laboratory.
Despite many years of study, this system continues to
offer new insight and has inspired many recent developments in
understanding topological phases of matter.
Given the huge impact of these ideas, it is natural to 
search for higher dimensional generalisations.

\para
A four-dimensional analog of the quantum Hall effect was suggested
some time ago by Zhang and Hu \cite{zhang,zhang2}. The original
proposal caused some excitement,
in large part because of the claim that the $d=3+1$ boundary
theory included a massless spin 2 particle, admittedly in conjunction with a tower of
 higher spin massless particles. Nervous about the violation of the spirit of
the Weinberg-Witten theorem, Elvang and Polchinski subsequently argued,
persuasively, that the boundary theory could not be a local quantum field theory
\cite{henriette}.
Our results for the  particular model of the 4d QHE considered here support this conclusion. 

\para
Nonetheless, the 4d quantum Hall effect (QHE)  contained enough
interesting structure to encourage further study, and there have been
a number of follow-up works exploring different aspects of this phase of matter 
\cite{fabinger,bernevig,chen,knair1,knair2,knair3,knair4}, as well as 
quantum Hall effects in higher dimensions \cite{eight,heckman}. 
 More recently, despite the very obvious dimensional limitations
 imposed by our Universe, there has even been a proposal to 
construct the 4d QHE in a cold atom system \cite{cold1,cold2}.

\para
There are some differences between our set-up below and the original
proposal of Zhang and Hu. In the latter, each particle moves on a
spatial ${\bf S}^4$ and carries an internal $SU(2)$ isospin index. This has the advantage that
one can introduce a background $SU(2)$ magnetic field, known as a
Yang-monopole, which preserves the full $SO(5)$ rotational invariance
\cite{yang}.
However, this also comes with some features which appear rather
unnatural.
In order to have a continuum limit, the $SU(2)$ isospin, $I$, of the
particle must scale with the radius $R$  of the ${\bf S}^4$ as $I\sim
R^2$.
This means that the original ``4d QHE'' is perhaps better thought of as  ``6d QHE", with the
particle moving on a phase space of volume $R^6$, which can be identified as 
${\bf CP}^3$ \cite{knair1}.

\para
In contrast, our particles will move in spatial ${\bf R}^4$.
We subject them to an Abelian magnetic field which breaks the  $SO(4)$ rotational symmetry. 
Nonetheless, the particles do carry internal degrees of freedom and
this dictates the physics. 

\para
Specifically, our particles arise as Yang-Mills instantons, viewed as
solitons in a $d=4+1$ $U(N)$ Yang-Mills theory\footnote{We note in passing that the Yang-monopole employed in
  \cite{zhang}  is really an instanton on ${\bf S}^4$. This means that
  instantons have never been far from the 4d quantum Hall effect; in
  our set-up they are dynamical, rather than providing a background field.}. For $U(1)$, these particles have no
internal degrees of freedom and the resulting ground states are the higher
dimensional generalisation of the Laughlin states.  For $N\geq 2$, the particles
carry both an internal isospin under $SU(N)$, and also a scale size.
This results in a higher dimensional generalisation of a class of 2d non-Abelian Hall states, 
first discovered by Blok and Wen \cite{bw}.

\subsubsection*{Relationship to the Nekrasov Partition Function}

In section \ref{partitionsec}, we compute the spectrum of excitations
above the quantum Hall ground state. We show that the 
resulting partition function is an object first introduced by Nekrasov 
in the context of ${\cal N}=2$ supersymmetric gauge theories in
$d=3+1$ 
dimensions \cite{nek}.

\para
We stress that  the Nekrasov partition function arises even though our
starting point is a many-body quantum mechanics involving purely
bosonic degrees of freedom.  
Morally speaking, the connection between the two can be 
traced to the importance of holomorphy in both supersymmetry and in
the lowest Landau level.

\para
Over the years, the  Nekrasov partition function has 
become a staple in the web of connections  linking gauge theories and
integrable systems  in various dimensions, and the original work has
been generalised in a myriad of different directions. Two of these 
generalisations will prove important in this work.

\para
First, one can define a Nekrasov partition function for $d=4+1$
supersymmetric theories on ${\bf R}^{3,1}\times {\bf S}^1$
\cite{nekok1}. This encourages a generalisation to include a 5d
Chern-Simons term at level $k$, 
and the resulting partition function was first computed in
\cite{yuji}; 
this is the version that will describe the excitations of the 4d
quantum Hall states, with $k$ related to the filling fraction of the state. 
The energies and other global charges 
of states in our model are encoded in the 
expansion of the Nekrasov partition function in powers of the equivariant
parameters which describe the $\Omega$-background and Coulomb branch
VEVs in the conventional setting of five dimensional supersymmetric
gauge theory. 

\para
For the purposes of understanding the ground state of our
model and its excitations, the standard formula for the
Nekrasov partition function as a sum over coloured partitions is not
particularly useful. Here we will 
make use of the standard fact that the Nekrasov partition function
counts holomorphic functions (or more generally 
holomorphic sections of an appropriate line bundle) 
on the instanton moduli space. The ADHM
construction gives a convenient realisation of
this manifold as a hyper-K\"{a}hler quotient or equivalently as the
Higgs branch of an auxiliary gauge theory. Finally, mirror
symmetry provides a dual description of the same manifold as the
Coulomb branch of yet another gauge theory which lives in three
spacetime dimensions.      
In recent work \cite{ami1,ami2,ami3},  this perspective has been used to provide 
alternative formulae for the Hilbert series of the
instanton moduli space.  
The equality of the resulting 
{\it Coulomb branch formulae} to the dual Higgs branch expression (i.e.
the sum over coloured partitions) is a highly non-trivial consequence of 3d
mirror symmetry. 

\para
In the following we will provide a derivation of
this Coulomb branch formula. We also  generalise this approach  to include the effects of a 5d
Chern-Simons term\footnote{As we will see in Section \ref{csaddsec}, the 5d Chern-Simons level corresponds to a background magnetic 
charge for a particular flavour symmetry in the 3d quiver gauge theory whose Coulomb branch coincides with the instanton moduli space.}. Happily the Coulomb branch formula provides exactly
the power series expansion of Nekrasov partition function we need to
analyse the ground state of our model and read off the spectrum of
low-lying modes.

\subsubsection*{Relationship to Quantum Hall  Matrix Models}

The story of this paper parallels our recent work on applying matrix 
model technology to the 2d QHE.

\para
A matrix model for the 2d Laughlin states was suggested
long ago by Polychronakos \cite{alexios} and subsequently extended to
non-Abelian quantum Hall states in \cite{us2}. Motivation for this description  originally came from  viewing the lowest Landau level as a non-commutative plane \cite{susskind}. 
It was subsequently realised that these matrix models describe the low-energy dynamics of vortices in a
$d=2+1$ dimensional $U(N)$ 
Yang-Mills-Higgs theory with Chern-Simons term at level $k$ 
\cite{unknown,us1}. Various properties of these matrix models have
been explored in great detail over the years; see, for example, 
\cite{hellvram,ks,ks2,hkk,cappelli,cappelli2,edge} for a selection of papers.

\para
In \cite{us3}, we computed the partition function of the quantum Hall
matrix models. This is equivalent to computing a corresponding Hilbert
series for the vortex moduli space. This partition function can
be written in closed form in terms of Kostka polynomials; 
the explicit combinatoric formula for the Kostka polynomials 
(originally due to Kirillov and Reshetikhin) is a direct analog of 
the ``Coulomb branch formula" for the Nekrasov partition function.
Importantly, we  could show that as the number of underlying particles becomes large, the vortex partition
function becomes the conformal block of the $U(N)_k$ chiral WZW model that 
lives at the boundary of the quantum Hall fluid.
In this way, the matrix model provides a direct link between the Chern-Simons
theory, the microscopic wavefunctions, and the boundary conformal field theory.

\para
The purpose of the present paper is to extend these ideas to $d=4+1$
dimensions. As we will show in the next section, the dynamics of the 
instantonic particles is described -- through the ADHM construction -- 
by a matrix model quantum mechanics. 
This will be our starting point for all that follows.

\section{A Quantum Hall Fluid of Instantons}

Our goal in this section is to describe the dynamics of Yang-Mills instantons in a background magnetic field, and determine the resulting quantum Hall wavefunctions governing their ground state. 

\subsection{The View from $d=4+1$ Dimensions}

Our starting point is $U(N)$ Yang-Mills theory in $d=4+1$ dimensions. We include also a single, real adjoint scalar field $\phi$. The action is then
\be S_{YM} = \int d^5x\ {\rm Tr}\,\left(-\frac{1}{2e^2}F_{\mu\nu}F^{\mu\nu} - \frac{1}{e^2}{\cal D}_\mu\phi{\cal D}^\mu\phi\right)\label{ymact}\ee
This theory enjoys a global $U(1)$ topological current, 
\be J_{\rm top}^\mu \sim \epsilon^{\mu\nu\rho\sigma\lambda}\,{\rm Tr}\,F_{\nu\rho}F_{\sigma\lambda}\nn\ee
under which solitonic excitations carry charge. These solitons are self-dual configurations, obeying
\be F_{\mu\nu} = {}^\star F_{\mu\nu} \ \ \ \ \ \mu,\nu=1,2,3,4\nn\ee
In the context of $d=3+1$ dimensional Yang-Mills, these objects are referred to as {\it instantons}. We will continue to use this name, even though it is less appropriate in $d=4+1$ dimensions where these are particles. Their mass is given by
\be M_{\rm inst} = \frac{8\pi^2}{e^2}\nn\ee
We will ultimately be interested in the dynamics of a large number of instantons, which we place  into a quantum Hall state. In order to do this, we will add a number of refinements to the action \eqn{ymact}. The first is a five-dimensional Chern-Simons term, which takes the form
\be S_{CS}  = \frac{k'}{24\pi^2} \int d^5x && \epsilon^{\mu\nu\rho\sigma\lambda}\,{\rm Tr}\,\left(A_\mu\partial_\nu A_\rho \partial_\sigma A_\lambda - \frac{3i}{2}A_\mu A_\nu A_\rho \partial_\sigma A_\lambda - \frac{3}{5} A_\mu A_\nu A_\rho A_\sigma A_\lambda\right) \nn\\ && \ \ \ +\ 
3d_{mnp}\phi^m\left(-\frac{1}{2}F_{\mu\nu}^n F^{p\,\mu\nu} - {\cal D}_\mu \phi^n{\cal D}^\nu \phi^p\right)
\label{csact}\ee
Here $d_{mnp} = \frac{1}{2}{\rm Tr}\,(T^m\{T^n,T^p\})$ is a symmetric tensor.

\para
The first line couples the topological current $J_{\rm top}$ to the gauge field, ensuring that instantons also carry non-Abelian electric charge. The requirement that action is invariant under large gauge transformations means that we must take $k'\in {\bf Z}$. The second line in \eqn{csact} means that the scalar field $\phi$ acts as an effective gauge coupling. The coefficients are fixed by the requirement of supersymmetry \cite{seib}. For our purposes, we're not interested in including any fermions so supersymmetry provides little motivation. However, including this coupling means that we can import various results about the dynamics of instantons in these theories \cite{kimyeong1,collie,kimyeong2}; we will describe these results in Section \ref{matrixsec}. 

\para
So far our particles are free to roam around. To coax them into a quantum Hall state, we include a coupling to a background magnetic field ${\cal A}_{\mu} J^{\mu}_{\rm top}$. We choose to work in symmetric gauge, with 
\be {\cal A}_\mu = \frac{B}{2}\bar{\eta}^3_{\mu\nu} x^\nu \nn\ee
with $B$ a constant and $\bar{\eta}^3_{\mu\nu}$ a 't Hooft matrix\footnote{The anti-self-dual 't Hooft matrices are give by
%
%
{\scriptsize
\be \bar{\eta}^1 = \left(\begin{array}{cccc} &  &  & -1 \\  &  & +1 &  \\  & -1 &  &  \\ +1 &  &  &   \end{array}\right)\ \ \ ,\ \ \ \bar{\eta}^2 = \left(\begin{array}{cccc}  &  & -1 &  \\  &  &  & -1 \\ +1 &  &  &  \\ & +1 &  &   \end{array}\right)\ \ \ ,\ \ \ \bar{\eta}^3 = \left(\begin{array}{cccc}  & +1 &  &  \\ -1 &  &  &  \\  &  &  & -1 \\  &  & +1 &   \end{array}\right)  \nn\ee} }. 
After an integration by parts, this can be written as a 3d Chern-Simons form, coupled to a constant background magnetic field $\omega = B \bar{\eta}^3$,
\be S_{\rm mag} = \int d^5x\  \omega_{\mu\nu}\,\epsilon^{\mu\nu\rho\sigma\lambda} {\rm Tr}\,\left(A_\rho \partial_\sigma A_\lambda - \frac{2i}{3}A_\rho A_\sigma A_\lambda\right)\label{magact}\ee
Such a term also arises in the ``K\"ahler Chern-Simons theory" of Nair and Schiff \cite{nairs1,nairs2}. This term breaks the spatial rotational symmetry to
\be SO(4) \cong SU(2)_L\times SU(2)_R \ \rightarrow\  SU(2)_L\times U(1)_R\nn\ee
This, then, is our goal: to understand the dynamics of instantons in the 5d theory with action 
\be S = S_{YM} + S_{CS} + S_{\rm mag}\nn\ee
As we will now see, there is an elegant description of the dynamics using the ADHM construction.

\subsection{The Matrix Model}\label{matrixsec}

The dynamics of $n$ instantons in Yang-Mills theory is beautifully described by a $U(n)$ gauged quantum mechanics, known as the ADHM matrix model \cite{adhm,douglas}. In its original formulation, the ADHM matrix model describes instantons in the Yang-Mills action \eqn{ymact}. However, as we explain below, it is simple to adapt the model to include the effects of  both $S_{CS}$ and $S_{\rm mag}$. 

\para

The ADHM data for $n$ interacting particles is built around a $U(n)$ matrix quantum mechanics. The fields are
\begin{itemize}
\item Two complex adjoint scalar fields, $Z$ and $\tilde{Z}$. These decompose as
\be Z = X^1 - iX^2\ \ \ {\rm and}\ \ \ \tilde{Z}=X^3+iX^4\nn\ee
Roughly speaking, the eigenvalues of each of these  $n\times n$
matrices parameterise the positions of the particles in the $x^\mu$,
$\mu=1,2,3,4$ directions. The ``roughly speaking" is because we will
often be dealing with situations in which the $X^\mu$ do not commute,
and so cannot be simultaneously diagonalised.

\item $N$ fundamental scalars $\varphi_i$ and $N$ anti-fundamental scalars $\tilde{\varphi}_i$, with $i=1,\ldots,N$. These endow our particles with a spin, transforming in some representation of $SU(N)$. This will be described in more detail below. 
\end{itemize}
The fields transform under a $U(n)$ gauge symmetry as
\be Z \rightarrow UZU^\dagger\ \ \ ,\ \ \ \tz\rightarrow U\tz U^\dagger\ \ \ ,\ \ \ \vp_i\rightarrow U\vp_i\ \ \ ,\ \ \ \tvp_i\rightarrow \tvp_i U^\dagger\label{gauge}\ee
with $U\in U(n)$.  

\para
These fields are required to obey two ADHM constraints, both of which are $n\times n$ matrix equations.  The first is real and is called the D-term; the second is complex and is called the F-term constraint (both names come from supersymmetric theories where these constraints are imposed by auxiliary fields),
\be \varphi_i\varphi_i^\dagger - \tvp_i^\dagger\tvp_i + [Z,Z^\dagger] + [\tz,\tz^\dagger] &=& 0\label{dterm0}\\ 
\vp_i\tvp_i -[Z,\tz] &=& 0\nn\ee
 Solutions to \eqn{dterm0}, subject to the identification \eqn{gauge}, define the instanton moduli space ${\cal  M}_{n,N}$. It has real dimension ${\rm dim}({\cal M}_{n,N}) = 4nN$.

\para
To impose the  symmetry \eqn{gauge}, we introduce a $U(n)$ gauge field $\alpha_0$. 
The dynamics of instantonic particles in the original Yang-Mills action \eqn{ymact} is governed by matrix quantum mechanics with configuration space ${\cal M}_{n,N}$. 
\be \tilde{S}_{YM} =  M_{\rm inst} \int dt &&   \,{\rm tr}\left(|{\cal D}_t Z|^2 + |{\cal D}_t \tilde{Z}|^2\right) + |{\cal D}_t\varphi_i|^2 + |{\cal D}_t\tilde{\varphi}_i|^2 \nn\\ &&\ \ \ -\ {\rm tr}(|[\sigma,Z]|^2 + |[\sigma,\tilde{Z}]^2 ) - \varphi_i\sigma^2 \varphi_i - \tilde{\varphi}_i\sigma^2\tilde{\varphi}_i^\dagger\nn\ee
Here ${\cal D}_tZ = \dot{Z} - i[\alpha_0,Z]$ and ${\cal D}_t\vp= \dot{\vp} - i\alpha_0\vp$ and ${\cal D}_t\tvp = \dot{\tvp} +i\tvp\alpha_0$, and the fields should be viewed subject to the constraints \eqn{dterm0}. In the second line, we have included one further, adjoint field $\sigma$; this plays no role here, but becomes important at the next step.

\para
We have two further terms in our 5d action. Each changes the ADHM matrix model in some way. The effect of the 5d Chern-Simons term is particularly simple; it introduces a worldline Chern-Simons term into the matrix model \cite{collie,kimyeong2},
\be \tilde{S}_{CS} =  -k' \int dt\ {\rm tr}(\alpha_0 +  \sigma)\nn\ee
Meanwhile, the effect of the 5d magnetic field \eqn{magact} was considered in \cite{andrew}. It adds a term to the action
\be \tilde{S}_{\rm mag} =  B\left(  i{\rm tr}(Z^\dagger {\cal D}_tZ + \tz^\dagger {\cal D}_t \tz) + i\vp^\dagger{\cal D}_t\vp + i({\cal D}_t\tvp_i)\tvp_i^\dagger \right) \nn\ee
We are left with the quantum mechanical matrix model $\tilde{S} = \tilde{S}_{YM} + \tilde{S}_{CS} + \tilde{S}_{\rm mag}$.

\subsubsection*{Projection to the Lowest Landau Level}

Quantum Hall physics arises in the limit of large magnetic field, in which the particles are restricted to lie in the lowest Landau level. Our interest  too lies in this regime, which is $B\gg M_{\rm inst}$. Operationally, one can reach this limit by rescaling lengths by $\sqrt{B}$, and subsequently sending $B\rightarrow \infty$. 

\para
It is straightforward to do this for the ADHM matrix model. In this limit, we lose the second order kinetic terms, leaving ourselves only with the first order terms,
\be {\cal L}_{\rm kin} &=&  i{\rm tr}(Z^\dagger {\cal D}_tZ + \tz^\dagger {\cal D}_t \tz) + i\vp^\dagger{\cal D}_t\vp + i({\cal D}_t\tvp_i)\tvp_i^\dagger - k'\,{\rm tr}\,\alpha_0  \label{kin} \ee
The fact that we have first order kinetic terms means that we're now viewing the instanton moduli space ${\cal M}_{n,N}$ as the phase space of the system, rather than the configuration space.

\para
There is one subtlety in taking the limit of large magnetic field. The original D-term constraint \eqn{dterm0} gets projected out in this limit. Instead, it is replaced by the Gauss law constraint from \eqn{kin} which reads
\be \varphi_i\varphi_i^\dagger - \tvp_i^\dagger\tvp_i + [Z,Z^\dagger] + [\tz,\tz^\dagger] &=& k'{\bf 1}_n\label{dterm}\ee
In contrast, the F-term constraint survives
\be \vp_i\tvp_i -[Z,\tz]  = 0\label{fterm}\ee
The Gauss law  \eqn{dterm} looks very much like the D-term for instantons in non-commutative space \cite{nekschwarz}. Here, however, it arises in a rather different context. As we will see in more detail shortly, the level $k'$ is telling us that wavefunctions should be sections of a line bundle over ${\cal M}_{n,N}$, rather than functions.

\subsubsection*{The Hamiltonian}

Our action \eqn{kin} contains only first order kinetic terms, and so defines the phase space of a system with vanishing Hamiltonian. To breathe some life into our system, we should define an energy function. To do so, note that the phase space enjoys a symmetry
\be G = U(1)_R\times SU(2)_L \times SU(N)\nn\ee
(If we set $k'=0$ then the phase space becomes singular and the $U(1)_R$ is enhanced to  $SU(2)_R$.) Under $G$, the fields transform as follows: 
$ (Z,\tz)$ sits in the representation  $({\bf 2},{\bf 1})_{+1}$;  $\vp$ in $({\bf 1},{\bf N})_{+1}$; and $\tvp$ in $({\bf 1},\bar{\bf N})_{+1}$.

\para
Later, we will study the spectrum of theory labelled by all these quantum numbers. For now, we want to pick one to define the Hamiltonian of the system. We choose the Hamiltonian to coincide with the $U(1)_R$ charge, 
%
%
\be H = \omega\left({\rm tr}(Z^\dagger Z + \tz^\dagger \tz) + |\varphi_i|^2  + |\tvp_i|^2\right)\label{ham}\ee
This has the interpretation of placing our system in a harmonic trap, which encourages the instantons towards the origin.

\para
This, then, is our matrix model for the 4d quantum Hall effect: a phase space  defined by \eqn{kin}, subject to \eqn{fterm}, with the Hamiltonian given by \eqn{ham}.  A model using similar field content, but differing in the details, was previously proposed in \cite{wrong1,wrong2}.

\subsection{Constructing the Hilbert Space}

The canonical commutation relations that arise from the Lagrangian \eqn{kin} are
\be [Z_{ab},Z^\dagger_{cd}] = [\tz_{ab},\tz^\dagger_{cd}] = \delta_{ad}\delta_{bc}\ \ \ ,\ \ \ [\vp_{ai},\vp^\dagger_{bj}] = [\tvp_{ai},\tvp^\dagger_{bj}] = \delta_{ab}\delta_{ij}\nn\ee
with all others vanishing. Here $a,b=1,\ldots,n$ label the $U(n)$ gauge indices of the various operators. These are, of course, the commutation relations of many creation and annihilation operators. We proceed in the usual manner by introducing the fiducial state $\ket{0}$ defined by
\be Z_{ab}\ket{0} = \tz_{ab}\ket{0} = \vp_{ai}\ket{0} = \tvp_{ai}\ket{0}=0\nn\ee
We then build a Hilbert space by acting with creation operators $Z^\dagger$, $\tz^\dagger$, $\vp^\dagger$ and $\tvp^\dagger$. However, not all states in this Hilbert space are physical: they must still obey the constraints \eqn{dterm} and \eqn{fterm}. We start with the Gauss law \eqn{dterm}. The traceless part of this condition is simply the requirement that physical states are $SU(n)$ gauge singlets. Meanwhile, the trace is 
\be \sum_{a=1}^k\sum_{i=1}^N\ \varphi_{ai}\varphi_{ai}^\dagger - \tvp^\dagger_{ai}\tvp_{ai} = k'n\ \ \ \Rightarrow\ \ \  \sum_{a=1}^k\sum_{i=1}^N\ \varphi_{ai}^\dagger\varphi_{ai} - \tvp^\dagger_{ai}\tvp_{ai} = (k'-N)n\ \ \ \  \label{trace}\ee
where the shift  $k'\rightarrow k'-N$ arises due to the normal
ordering and was seen in earlier studies \cite{alexios,us1,us2}. In the 
matrix model for the 
2d QHE, this was traced to a one-loop shift in the 3d Chern-Simons
term and 
it seems plausible that a similar interpretation holds here. 
We define
\be k = k'-N\nn\ee
The result \eqn{trace} tells  us that we must have an excess of $kn$  
$\vp$ excitations over $\tvp$ excitations. 

\para
Finally, we have the F-term constraint \eqn{fterm}. One can treat this
as an operator equation which must annihilate any physical
state. However, there is an equivalent interpretation that is
sometimes more useful:  in any gauge invariant state, we may always
commute  $Z^\dagger$ and $\tz^\dagger$ at the cost of introducing
$(\vp_i\tvp_i)^\dagger$ insertions. This means that we can restrict
ourselves to representative states in which all $Z$ operators precede
the $\tz$, while we allow any number of $\vp$ and $\tvp$ operators.
This then 
leaves us with the following $SU(n)$ singlet operators to work with: baryons,
\be B \sim \epsilon^{a_1\ldots a_n} (Z^{l_1}\tz^{m_1}\varphi_{i_1})_{a_1}\ldots(Z^{l_n}\tz^{m_n}\varphi_{i_n})_{a_n}\label{baryon}\ee
and
\be \tilde{B} \sim \epsilon^{a_1\ldots a_n} (\tilde{\varphi}_{i_1}Z^{l_1}\tz^{m_1})_{a_1}\ldots(\tilde{\varphi}_{i_n}Z^{l_n}\tz^{m_n})_{a_n}\nn\ee
mesons,
\be M \sim \tilde{\vp}_i Z^l \tz^m\vp_j\label{meson}\ee
and ``glueballs"
\be G \sim {\rm tr}\,Z^l\tz^m\label{glueball}\ee
All holomorphic gauge invariant operators in the theory can be obtained
as products of these generators\footnote{Note  that the resulting chiral
  ring has additional relations which are
  always present for finite dimensional matrices. However, in the
  limit of large instanton number (for $k=0$), these constraints are
  relaxed and the chiral ring is freely generated.}.   
The Hamiltonian $H$ simply counts the number of creation operators of
each type. The condition \eqn{trace} tells us that we must have an
excess of 
$\varphi$ over $\tvp$ operators and the only operators above with this
property are the baryons $B$.  As we now explain, this results in an 
interesting pattern of quantum numbers for the ground state, depending 
on our choice of the integers $N$, $n$ and $k$.

\subsection{Abelian Quantum Hall States}\label{abeliansec}

We start by describing the Abelian quantum Hall states that arise when
we take $N=1$. We will describe the non-Abelian states in 
Section \ref{nonabsec}. 

\subsubsection*{A Single Particle}

For a single $n=1$ particle, the matrix model is trivial. First, the $\vp$ and $\tvp$ operators are fixed completely by the D- and F-term constraints \eqn{dterm} and \eqn{fterm}. All physical states in the Hilbert space are dressed by $\vp^{\dagger\,k}\ket{0}$. No $\tvp$ excitations are allowed.

\para
All the dynamics lies in the position operators $Z$ and $\tilde{Z}$. These are neutral under the $U(1)$ gauge group and are governed by the free action
\be S = \int dt\ i Z^\dagger \partial_tZ + i \tilde{Z}^\dagger \partial_tZ - \omega (|Z|^2 + |\tilde{Z}|^2)\nn\ee
This describes a single particle, moving in ${\bf R}^4$, projected to
the lowest Landau level by a strong background magnetic field
$F_{\mu\nu}\sim \bar{\eta}^3_{\mu\nu}$.  (We have set the strength of
this magnetic field to 1; this, in turn, means that the coordinates on
${\bf R}^4$ are dimensionless.)  Note that the system can
be viewed as two, decoupled planar systems, one with a magnetic field
in the $1-2$ plane and the other with a magnetic field in the $3-4$
plane. This magnetic field breaks the $SO(4)$ rotational symmetry of
${\bf R}^4$ to $SU(2)_L \times U(1)_R$.

\para
The potential term, proportional to $\omega$, acts as a harmonic trap, encouraging the particle to lie near the origin. The single particle eigenstates are given by
\be \ket{l,m} = Z^{\dagger\,l}\tilde{Z}^{\dagger\,m}\vp^{\dagger\,k}\ket{0}\nn\ee
These have energy
\be E_{l,m} = \omega(l+m)\nn\ee
The ground state is $\ket{0,0}$. 
States with higher energy sit further from the origin and, correspondingly, lie in higher dimensional representations of the rotation symmetry $SU(2)_L$. Specifically, the states $\ket{l,m}$ have degeneracy $l+m+1$. These sit in a multiplet of $SU(2)$ with spin
\be s = \frac{1}{2}(l+m)\nn\ee
This single particle intuition will be useful when discussing the many particle case.

\subsubsection*{Many Particles}

The $U(n)$ matrix model (again, with $N=1$) describes many, interacting particles, all projected to the lowest Landau level in ${\bf R}^4$. We first focus on the ground state. The D-term constraint \eqn{trace} tells us that all physical states must have $kn$ $\vp$ excitations, while the Hamiltonian \eqn{ham} tells us that the ground state has the fewest excitations possible. This means that that ground state is built from the baryon operator \eqn{baryon} that contains the fewest creation operators. 

\para
Because the $\vp$ are commuting operators, the baryon \eqn{baryon} is only non-vanishing if we put a distinct $Z^l\tilde{Z}^m$ in each slot. The baryon with the lowest energy is then
\be B  = \epsilon^{a_1\ldots a_n} \varphi_{a_1} (Z\vp)_{a_2}(\tz\vp)_{a_3} (Z^2\vp)_{a_4}\ldots \ldots(Z^{l_n}\tz^{m_n}\varphi)_{a_n}\label{lowestb}\ee
One can think about this operator rather like filling atomic energy levels. In the present case, these are the  $SU(2)_L$ multiplets of a single particle. The operator $B$ is unique only when a shell is fully-filled. This happens when the number of particles takes the form
\be n = (s+1 )(2s+1)\ \ \ s\in \frac{1}{2}{\bf Z}\label{triangle}\ee
In this case, all $SU(2)$ representations up to spin $s$ are filled. Only then is the resulting baryon operator an $SU(2)_L$ singlet. 

\para
When $n$ takes the form \eqn{triangle}, the ground state of the matrix model is given by
\be \ket{\Omega}  = B^{\dagger\,k}\ket{0}\label{abelianground}\ee
and has energy
\be E_0 = \frac{\omega k}{3}(s+1)(2s+1)(4s+3) \sim \frac{\omega k}{3} (2n)^{3/2} \label{energy1}\ee
We will describe the physical interpretation of this state below.

\para
When the number of particles $n$ does not take the form \eqn{triangle} there is no unique ground state. Instead, the baryon operator \eqn{lowestb} sits in a representation of $SU(2)_L$ which, typically, is not an irrep. For example, when $n=(s+1)(2s+1) + 1$, the $B$ sits in the spin $s+1$ representation of $SU(2)_L$. When $n=(s+1)(2s+1)+2$, the baryons sits in the anti-symmetrised representations of $2s+1 \otimes 2s+1$ (i.e. every alternate representation in the decomposition). The ground state \eqn{abelianground} then sits in the symmetrised representations of $k$ products of the representation of $B$. In general, we are left with a state which transforms in a large slew of different irreps.

\subsubsection*{Wavefunctions}

In order to illuminate the physical meaning of the ground state \eqn{abelianground}, it would be useful to write it in terms of a wavefunction $\psi(z_a,\tilde{z}_a)$ where $(z_a,\tilde{z}_a)$ describes the position in ${\bf R}^4$ of the $a^{\rm th}$ particle. A number of techniques have been developed for converting states in these matrix models into wavefunctions for point-like particles \cite{hellvram,ks,ks2}.

\para
Here we work with the coherent state representation. We sketch only the bare details; more information can be found in \cite{ks}. These coherent states $\ket{Z,\tilde{Z},\vp,\tvp}$ obey
\be \hat{Z}\ket{Z,\tilde{Z},\vp,\tvp} = Z\ket{Z,\tilde{Z},\vp,\tvp}\nn\ee
where, for once, we've introduced a hat to distinguish the quantum matrix operator $\hat{Z}$ from the classical matrix $Z$. Similar equations hold for $\tilde{Z}$, $\vp$ and $\tvp$.
Suppose one subsequently diagonalises both $Z= VDV^{-1}$ and $\tilde{Z} = V\tilde{D}V^{-1}$ where $D={\rm diag}(z_1,\ldots, z_n)$ and $\tilde{D} = {\rm diag}(\tilde{z}_1,\ldots,\tilde{z}_n)$, and constructs the wavefunction $\psi(z,\tilde{z}) = \bra{z,\tilde{z}}\Omega\rangle$ -- one might imagine this is possible as a consequence of the constraint $[Z,\tilde{Z}] = 0$. One typically expects that such a change of variables induces a Jacobian which can alter the nature of the resulting state importantly, but for simplicity we will focus only on the $k$-dependent part of the wavefunction.\footnote{It is worth recalling what the Jacobian does in matrix models for the 2d quantum Hall effect \cite{alexios, ks}. In this case, the Jacobian takes the form $\prod_{b<c} |z_b-z_c|^2$, which naturally factorises into $\mbox{holomorphic}\times \mbox{anti-holomorphic}$. The holomorphic part is then absorbed into the wavefunction $\psi$, where it has the effect of shifting the filling fraction of the Laughlin state from $k\rightarrow k+1$. Similarly, the anti-holomorphic part is absorbed into $\phi$.}


\para
As we have seen, life is simplest when the number of particles takes the form \eqn{triangle}. The asymptotic form of the wavefunctions, $|z_i-z_j|, |\tilde{z}_i-\tilde{z}_j| \gg 1$, contains the universal information describing the state. The ground  state \eqn{abelianground} has asymptotic behaviour
\be \bra{z,\tilde{z}}B^{\dagger\,k}\ket{0}  = \psi(z,\tilde{z}) \rightarrow \tilde{\psi}^k_0(z,\tilde{z})\,e^{-\frac{1}{4}\sum_a|z_a|^2 + |\tilde{z}_a|^2}\label{wf}\ee
where $\tilde{\psi}_0(z,\tilde{z})$ is a Slater determinant which describes a fully filled Landau level in four dimensions, 
\be  \tilde{\psi}_0(z_i,\tilde{z}_i) \equiv \left|\begin{array}{cccccccccc} 1 & z_1 & \tilde{z}_1 & z_1^2 & z_1\tilde{z}_1 & \tilde{z}_1^2 & z_1^3 & \ldots & z_1 \tilde{z}_1^{2s-1} & \tilde{z}_1^{2s} \\ 1 & z_2 & \tilde{z}_2 & z_2^2 & z_2\tilde{z}_2 & \tilde{z}_2^2 & z_2^3 & \ldots & z_2 \tilde{z}_2^{2s-1} & \tilde{z}_2^{2s}  \\ 
\vdots & &&&&&&&& \vdots \\
1 & z_n & \tilde{z}_n & z_n^2 & z_n\tilde{z}_n & \tilde{z}_n^2 & z_n^3 & \ldots & z_n \tilde{z}_n^{2s-1} & \tilde{z}_n^{2s} 
\end{array}\right|\label{slater}\ee
For $k=1$, this state  can be thought of as an incompressible fluid, filling a ball of radius $R^2 \sim 4s$. The volume of this ball is $V=\frac{1}{2}\pi^2 R^4 \sim 4\pi^2 n$.

\para
For $k>1$, the wavefunction $\tilde{\psi}_0^k(z,\tilde{z})$ can be viewed as the higher dimensional version of the Laughlin state. These states were also constructed in \cite{liwu} where the authors pointed out that they satisfy a quaternionic version of analyticity. 

\para
In two dimensions, the Slater determinant -- and, correspondingly, the Laughlin wavefunctions -- famously take the product form $\prod_{a<b}(z_a-z_b)^k$. This, of course, has the property that it vanishes whenever two particles coincide. In four dimensions, the Slater determinant \eqn{slater} also vanishes whenever two particles coincide, but now this requires two complex conditions
\be z_a=z_b \ \ \ {\rm and}\ \ \ \tilde{z}_a=\tilde{z}_b\nn\ee
The fact that the wavefunction vanishes only when two conditions are met means  that it cannot be written as a single product of holomorphic factors of complex variables.

\subsection{Excited States}

We now turn to the spectrum of excited states. At first glance, it appears that there are three, distinct  low-energy excitations above $\ket{\Omega}$. These arise by acting with  the mesons \eqn{meson} or  glueball \eqn{glueball} operators, or by increasing the $Z^l\tilde{Z}^m$ factors inside the baryon operators \eqn{baryon}.

\para
However, not all of these give rise to independent, physical excitations. First, it turns out that the F-term constraint \eqn{fterm} excludes all meson operators \eqn{meson} when $N=1$. This is even true at the classical level (see Proposition 2.8 of \cite{nakajima}).

\para
This leaves us with only the baryonic and glueball excitations. These are not independent. The full relationship between them is complicated. We will describe the spectrum of excited states in Section \ref{partitionsec} where we compute the partition function of this matrix model. 

\para
For now, we make one comment about the spectrum of excitations. We start with the $SU(2)_L$ singlet ground state that arises when the number of particles takes the form \eqn{triangle}. There is a large number of minimal energy excitations, with $E = E_0+\omega$, that arise by promoting one particle from the filled spin-$s$ shell of the baryon \eqn{lowestb} to the unfilled spin-$(s+1)$ shell. The degeneracy of these excitations is given by $(2s+1)(2s+3)$ and they decompose into representations $1\oplus 2 \oplus \ldots 2s+1$. The existence of such a large number of low-lying excitations is the crux of the argument in \cite{henriette} that there is no local $d=3+1$ boundary theory for the 4d quantum Hall effect. We will describe these boundary excitations in more detail in Section \ref{partitionsec}.

\subsection{Non-Abelian Quantum Hall States}\label{nonabsec}

Let's now turn to the non-Abelian states, based on instantons in a $U(N)$ gauge theory. The matrix model fields and constraints were described in \eqn{gauge}, \eqn{dterm} and \eqn{fterm}.

\subsubsection*{A Single Particle with Spin}

To start, we can build some intuition for this matrix model by considering a single particle. The matrix model is based around a $U(1)$ gauge group so the adjoint fields $Z$ and $\tilde{Z}$ now decouple, and describe a particle on ${\bf R}^4$ sitting in the lowest Landau level. We're left with the internal degrees of freedom $\varphi_i$ and $\tilde{\varphi}_i$ which must obey
\be \sum_{i=1}^N \varphi_i^\dagger \varphi_i - \tilde{\varphi}_i^\dagger \tilde{\varphi}_i = k \ \ \ {\rm and}\ \ \ \sum_{i=1}^N\tilde{\varphi}_i\varphi_i = 0\label{u1constraints}\ee
This describes the  cotangent bundle $T^\star {\bf CP}^{N-1}$. Endowed with  the first order kinetic terms \eqn{kin}, this is the phase space of the internal degrees of freedom.

\para
The cotangent bundle $T^\star {\cal M}$ usually appears as the phase space for a particle moving on ${\cal M}$. This is not the correct interpretation in the present case. Instead, the symplectic form arising from \eqn{kin} means that, upon quantising $T^\star {\bf CP}^{N-1}$, one finds that the particle carries an internal degree of freedom which we will refer to as ``spin", transforming in a representation of $SU(N)$. 

\para
To describe this, we first look at the ground state of the Hamiltonian \eqn{ham}, which is now given by
\be |\Omega_{i_1\ldots i_k}\rangle = \prod_{m=1}^{k} \varphi_{i_m}^\dagger\ket{0}\nn\ee
This describes the Hilbert space of a particle which sits in the $k^{\rm th}$ symmetric representation of $SU(N)$. In terms of Young diagrams, this is $k$ boxes 
\be \yng(4)\nn\ee
For example, when $k=1$ the particle has $N$ internal states and transforms in the fundamental representation $\yng(1)$ of $SU(N)$. 

\para
Even a single particle has a  spectrum of excited states. These come from acting with $\tilde{\varphi}$. Because of the D-term constraint \eqn{u1constraints}, we must have an equal number of $\vp$ and $\tvp$ excitations. The excited states at level $p$ take the form
\be \prod_{m=1}^p (\tvp_{j_m} \vp_{l_m})^\dagger \ket{\Omega_{i_1\ldots i_k}}\label{unphysical}\ee
for some collection of flavour indices $\{j_m,l_m|m=1,\ldots,p\}$. The number of such states is given by ${{N+p-1}\choose{N-1}} {{N+p+k-1}\choose{N-1}}$. However, not all of these obey the F-term constraint $\sum_i\tvp_i\vp_i\ket{\rm phys}=0$. Acting with this gives ${N+p-2\choose N-1}{N+p+k-2\choose N-1}$ linearly independent constraints. Thus, the dimension of the Hilbert space at level $p$ (where $p=0$ is the ground state) is given by
\be {\rm dim}{\cal H}_p ={N+p-1\choose N-1}{N+p+k-1\choose N-1}-  {N+p-2\choose N-1}{N+p+k-2\choose N-1}\nn\ee
These form a single, irreducible representation of $SU(N)$ at level $p$. Indeed, the states \eqn{unphysical} sit in the representations ${\rm Sym}^p(\bar{\bf N}) \otimes {\rm Sym}^{p+k}({\bf N})$. The F-term constraint projects onto the highest dimension representation which, in terms of Young diagrams, looks like this: 
\be \yng(7,3,3,3,3)\nn\ee
Here, the first row has $2p+k$ boxes and all other rows have $p$ boxes, and there are $N-1$ rows in total. The diagram shown above is the representation of the $p=3$ excited state for a particle in  $SU(6)$ with $k=1$. Note that, for $k=0$, this coincides with the holomorphic functions on the instanton moduli space described in \cite{ami67}. We will see in section \ref{nicesec} how the above projection can be proven using the Littlewood-Richardson rule at the level of the partition function.

\para
The upshot of this discussion is that the moduli space of a single instanton describes a single particle with a tower of excited states, each of which sits in increasingly higher dimensional representations of the $SU(N)$ flavour symmetry. We can view the particle as containing an internal degree of freedom transforming in the adjoint of $SU(N)$; the excitations  then carry an extra  ${\rm Sym}^p({\bf Adj})$ group structure, which dresses the $k^{\rm th}$ symmetric representation of the ground state. 

\para
Of course, from the perspective of the instanton equations we expect that this tower  corresponds to the instanton growing in size. This interpretation isn't obviously apparent in the description of the Hilbert space presented above.

\subsubsection*{The Non-Abelian Ground State}

Let us now discuss the ground state of the non-Abelian matrix model describing many instantons. Because it costs energy to excite the meson fields, the ground state contains only $\vp$ excitations and no $\tvp$ excitations. This means that we are describing a ground state of $n$ particles, each of which transforms in the $k^{\rm th}$ symmetric representation of $SU(N)$. 

\para
For a generic number of instantons, this ground state is neither a singlet of $SU(N)$ nor $SU(2)_L$. However, nice things happen when the number of instantons is given by
\be n = N(s+1)(2s+1)\ \ \ \  s\in \frac{1}{2}{\bf Z}\label{magicn}\ee
In this case, there is a unique ground state. To describe this, we first introduce the $SU(N)$ baryon operators. These take the form
\be B(l,m)_{a_1\ldots a_N} = \epsilon^{i_1\ldots i_N} (Z^l\tilde{Z}^m \varphi)_{i_1\,a_1}\ldots  (Z^l\tilde{Z}^m \varphi)_{i_N\,a_N}\label{minib}\ee
This is an $SU(N)$ singlet and  transforms in the $N^{\rm th}$ anti-symmetric representation of the $U(n)$ gauge symmetry. The ground state is then formed by building up consecutive baryons which minimize the value of $l+m$. This means that we first act with $B^\dagger(0,0)$, followed by $B^\dagger(1,0)$ and $B^\dagger (0,1)$, and so on. When the number of instantons takes the form \eqn{magicn} there we have precisely the right number to fill the $SU(2)_L$ shells, up to spin $s$. The resulting singlet state is given by
\be \ket{\Omega}  = \left[\epsilon^{a_1\ldots a_n} B^\dagger(0,0)_{a_1\ldots a_N}B^\dagger(1,0)_{a_{N+1}\ldots a_{2N}}\ldots B^\dagger(0,2s)_{a_{n-N+1}\ldots a_n}\right]^k\ket{0}\label{nonabstate}\ee
This state has energy $NE_0$, where $E_0$ is the ground state energy of the Abelian quantum Hall state  \eqn{energy1}. 

\para
The state \eqn{nonabstate} is the four-dimensional generalisation of the Blok-Wen states \cite{bw}. In the 2d QHE, the Blok-Wen states  are a particularly simple class of  non-Abelian quantum Hall states. They describe particles which each carry an internal spin, transforming in the $k^{\rm th}$ symmetric representation of $SU(N)$. They have the special property that the associated boundary $d=1+1$ conformal field theory is the $SU(N)_k$ WZW model. The 2d Blok-Wen states also arise as the ground state of a matrix model associated to the vortex moduli space \cite{us2}. Here we see their 4d counterparts emerging from the ADHM matrix model.

\para
The excitations above the ground state arise by acting with meson \eqn{meson} and glueball \eqn{glueball} operators, as well as by changing the occupied shells inside the $k$ baryon operators. As in the Abelian case, there are complicated relations between these and we postpone a detailed discussion to Section \ref{partitionsec} where we compute the partition function. One noticeable difference from the Abelian case is that now the meson operators give new degrees of freedom, characterised by their quantum numbers under $SU(N)$. 

\subsubsection*{Non-Abelian Wavefunctions}

We can, once again, translate these states into wavefunctions. Now there is an extra subtlety, because, as we have seen, each particle carries an internal spin degree of freedom. In the ground state these transform in the $k^{\rm th}$ symmetric representation of $SU(N)$.  It will be useful to work  through some examples.

\para
\underline{$SU(2)$, $k=1$:} 

\para
In this case, each instanton carries spin $\frac{1}{2}$ under the $SU(2)$ global symmetry. There are two states: $\up$ and $\down$. The wavefunctions therefore depend on both the positions $(z,\tilde{z})$ and the spin states. The way to interpret matrix model states as wavefunctions with spin was described in detail in \cite{us2}. Omitting the overall exponential \eqn{wf}, the holomorphic part of the ground state wavefunction is given by $\tilde{\psi}(z,\tilde{z},{\rm spin}) = \Phi(z,\tilde{z},{\rm spin})$, where
\be \Phi(z,\tilde{z},{\rm spin}) &=& \epsilon^{a_1\ldots a_n}(1_{a_1}1_{a_2})\,(z_{a_3}z_{a_4})(\tilde{z}_{a_5}\tilde{z}_{a_6})\,(z^2_{a_7}z^2_{a_{8}})(z_{a_{9}}\tilde{z}_{a_9}z_{a_{10}}\tilde{z}_{a_{10}}) (\tilde{z}_{a_{11}}^2\tilde{z}_{a_{12}}^2)\ldots \nn\\ &&\ \ \ \ \ \ \ \ \ \ \ \  \ldots (\tilde{z}^{2s}_{a_{n-1}}\tilde{z}^{2s}_{a_n}) \times\ \Big[\ket{\uparrow_{a_1}}\,\ket{\downarrow_{a_2}}\,\ket{\uparrow_{a_3}}\,\ket{\downarrow_{a_4}} \ldots \ket{\uparrow_{a_{n-1}}}\,\ket{\downarrow_{a_n}}   \Big]\nn\ee
Here, each bracket corresponds to an $SU(2)$ baryon of the form \eqn{minib}. The anti-symmetrisation ensures that the spins are arranged as $\ket{\uparrow_{a_1}}\ket{\downarrow_{a_2}}- \ket{\downarrow_{a_1}}\ket{\uparrow_{a_2}}$, which is an $SU(2)$ singlet.

\para
\underline{$SU(2)$, $k>1$:}

\para
For $k>1$, each instanton sits in the $k^{\rm th}$ symmetric representation which, for $SU(2)$, means spin $k/2$. Each instanton is now labelled by an internal spin state $\ket{\sigma}$ with $\sigma = -k/2,\ldots, +k/2$. The ground state wavefunction is given by
\be \tilde{\psi}(z,\tilde{z},\sigma) = {\cal P}\Big[ \otimes^k\Phi(z,\tilde{z},{\rm spin})\Big]\label{bwsu2}\ee
For each particle, the spin-$\frac{1}{2}$ state $\up$ or $\down$ appears $k$-times in tensor product. The symbol ${\cal P}$ tells us that we should project this onto the fully symmetrised, spin $k/2$ representation. 

\para
For example, when $k=2$, the product $\Phi\otimes \Phi$ will have two spin-$\frac{1}{2}$ states for each particle. We interpret these as spin-1 states using
\be \up\up = \ket{1}\ \ \ ,\ \ \ \up\down = \ket{0}\ \ \ ,\ \ \ \down\down = \ket{-1}\nn\ee
For any $k$, the wavefunction \eqn{bwsu2} is a singlet under the internal  $SU(2)$. (It is also a singlet under the rotational $SU(2)_L$.) This fact was proven in \cite{us2}. 

\para
\underline{$SU(N)$:}

\para
The story for $SU(N)$ is similar to that described above. We start with $k=1$, where each instanton carries an internal degree of freedom in the fundamental of $SU(N)$. We write this as $\ket{\sigma}$, with $\sigma=1,\ldots N$. Given $N$ particles, we can form a spin singlet from the baryon
\be b_{a_1\ldots a_N} = \epsilon^{\sigma_{a_1}\ldots\sigma_{a_N}}\ket{\sigma_{a_1}}\ldots \ket{\sigma_{a_N}}\nn\ee
The wavefunction corresponding to \eqn{nonabstate} is then $\tilde{\psi} = \Phi(z,z,\sigma)$ where
\be \Phi_{N}(z,\tilde{z},\sigma) &=& \epsilon^{a_1\ldots a_n}(1_{a_1}\ldots 1_{a_N})\,(z_{a_N+1}\ldots z_{a_{2N}})(\tilde{z}_{a_{2N+1}} \ldots \tilde{z}_{a_{3N}})\,(z^2_{a_{3N+1}}\ldots z^2_{a_{4N}})\ldots \nn\\ &&\ \ \   \ldots (z_{a_{4N+1}}\tilde{z}_{a_{4N+1}}\ldots z_{a_{5N}}\tilde{z}_{a_{5N}})   (\tilde{z}^{2s}_{a_{n-N+1}}\ldots \tilde{z}^{2s}_{a_n}) \times\ \Big[ b_{a_1\ldots a_N}\ldots b_{a_{n-N+1}\ldots {a_n}}\Big]\nn\ee
This state has clustering at order $N$: the wavefunction remains non-zero if the positions of up to $N$ particles coincide. The wavefunction vanishes if $N+1$ or more particles coincide. 

\para
For $k>1$, each particle carries an internal spin, transforming in the $k^{\rm th}$ symmetric representation of $SU(N)$. The ground state \eqn{nonabstate} is given by
\be \tilde{\psi}(z,\tilde{z},\sigma) = {\cal P}\Big[ \Phi^k(z,\tilde{z},\sigma)\Big]\nn\ee
where, once again, ${\cal P}$ projects onto the symmetric product of spin states. It was shown in \cite{us2} that this state is an $SU(N)$ spin singlet.

\section{The Partition Function}\label{partitionsec}

Our goal in this section is to compute the partition function of the matrix model. Our partition function will depend on the fugacities for the Cartan subalgebra of the full symmetry group  $G = U(1)_R\times SU(2)_L \times SU(N)$. Although the physical Hilbert space contains only $SU(n)$ singlets (of fixed $U(1)$ charge) under the $U(n)$ gauge symmetry, it will prove useful in our intermediate calculations to also introduce fugacities for this gauge symmetry. We denote the fugacities and quantum numbers of the various Cartan elements as
\begin{center}
\begin{tabular}{c|cccc}
 & $U(1)_R$ & $SU(2)_L$ & $SU(N)$ & $U(n)$  \\ 
 \hline
Quantum number & $ \Delta$ & $J$ &  $j_i$ & - \\  
Fugacity & $q$ & $z$ & $x^i$ & $w^a$
\end{tabular}
\end{center}
Note that we have one redundant fugacity in $SU(N)$, which means that we can decide to fix $\prod_i x_i = 1$ if we wish\footnote{Alternatively, one can keep the product, which then simply tracks the fixed $U(1)$ gauge charge of the system.}.
The Hamiltonian \eqn{ham} is proportional to  the $U(1)_R$ charge, so
the energy of any state is $E=\omega \Delta$.

\para
We  compute the partition function
\be {\cal Z}(q,z,x_i) = {\rm Tr}\,q^\Delta z^{J}\prod_{i=1}^N x_i^{j_i}\label{part}\ee
where the trace is over all states in the Hilbert space. This means
that they obey both the Gauss law constraint \eqn{dterm} and the
F-term constraint \eqn{fterm}. Note that, for zero Chern-Simons level, 
our states are in one-to-one
correspondence with 
holomorphic functions on the instanton moduli space. When $k\neq 0$
the resulting baryonic states correspond instead to holomorphic
sections. 
This means that the partition function for our theory 
coincides with the Hilbert series, a fact we will use in Section \ref{amisec}.

\subsection{The Nekrasov Partition Function}\label{nekrasov-sec}

We will first write down an expression for the partition function ${\cal Z}$ in integral form. To do this, we enumerate all gauge-variant observables subject to the F-term constraint \eqn{fterm}, and subsequently restrict to the gauge invariant sector. The integral has a number of pieces:
\begin{itemize}
\item The adjoint fields $Z$ and $\tilde{Z}$ lie in the adjoint of the gauge group, and so carry quantum numbers $w_aw_b^{-1}$ for some $a\neq b$. They also form a doublet of $SU(2)_L$. They contribute factors to the partition function given by
\be {\cal Z}_Z = \prod_{a,b=1}^n\frac{1}{1-qzw_a/w_b}\ \ \ ,\ \ \  {\cal Z}_{\tilde{Z}} = \prod_{a,b=1}^N\frac{1}{1-qw_a/zw_b}\nn\ee
\item The (anti)-fundamental fields $\vp$ and $\tvp$ sit in the $({\bf N},{\bf n})$ and $(\bar{\bf N},\bar{\bf n})$ representations of $SU(N)\times U(n)$ respectively. Their contribution to the partition function is given by
\be {\cal Z}_\vp = \prod_{a=1}^n\prod_{i=1}^N \frac{1}{1-qx_iw_a} \ \ \ ,\ \ \ {\cal Z}_{\tvp} = \prod_{a=1}^n\prod_{i=1}^N \frac{1}{1-q/x_iw_a}\nn\ee
\item The F-term constraint \eqn{fterm} can be implemented by including the factor
\be {\cal Z}_{F} = \prod_{a,b=1}^n\left(1-q^2\frac{w_a}{w_b}\right)\label{fz}\ee
which subtracts from the partition function the appropriate terms which carry the quantum numbers of the constraint.
 \end{itemize}
We then impose $U(n)$ gauge invariance by a suitable contour integral. Including the Haar measure on the group manifold $U(n)$, the partition function is given by the Molien integral
\be {\cal Z}(q,z,x_i) = \frac{1}{n!}\left(\prod_{a=1}^n \frac{1}{2\pi i}\oint \frac{dw_a}{w_a^{k+1}}\right)\prod_{b\neq c}\left(1-\frac{w_b}{w_c}\right){\cal Z}_Z{\cal Z}_{\tilde{Z}}{\cal Z}_\vp{\cal Z}_{\tvp}{\cal Z}_F\label{z}\ee
The integration contours are taken around $|w_a|=1$, while the other fugacities are understood to take values $|q|,|z| \ll 1$ and $|x_i|\sim 1$.  The integrals over $w_a$ project onto $SU(n)$ gauge singlet states while the extra factor of $\prod w_a^{-k}$ ensures that the partition function counts only those states that have $U(1)\subset U(n)$ charge $k$. This captures the role of the level $k$ in the Gauss law constraint \eqn{dterm}.

\para
The partition function \eqn{z} is a variant of the Nekrasov partition
function \cite{nek}. More, precisely, this is the Nekrasov partition
function for 5d, ${\cal N}=1$ $SU(N)$ super-Yang-Mills compactified on
${\bf R}^4\times {\bf S}^1$ \cite{nekok1} in an 
$\Omega$ background. In this context, the level  
$k$ is associated to a 5d Chern-Simons term, as first discussed in \cite{yuji}. 
The fugacities $x_{i}$ for the global $U(N)$ symmetry are identified
with the Coulomb branch parameters of the 5d SUSY theory while the
fugacities, $z$ and $q$ of the two $SU(2)$ 
rotation symmetries of $\mathbb{R}^{4}$ are identified with the
parameters of the $\Omega$-background.  
\para
The integral is not the usual form of the Nekrasov partition function,
which is typically written as a sum over coloured 
Young tableaux.  (The partition function can be found written  
in this integral form for $N=1$ and $k=0$ in, for example,
\cite{nekok,koreans}.) It is instructive to see how the more familiar
expression arises from the pole structure of the integral when
$k=0$. The poles sitting inside the contour arise whenever $w_a=qx_i$
or $w_b=qz w_a$ or $w_b = qw_a/z$. (The pole from 
${\cal Z}_\vp$ when $\omega_a = 1/qx_i$ sits outside the unit circle,
and the pole at $w=0$ is killed by the fact that ${\cal Z}_{\tvp}$
vanishes there.)  The set of poles, up to $S_n$ permutation, is then
characterised in the following 
way: we divide $w_a$ up into $N$ groups, each of size $n_i$, with
$\sum_i n_i = n$. In each group, one element gets its pole from $w_a =
qx_i$, while the 
others build up from this, either through $w_b=qzw_a$ or
$w_b=qw_a/z$. The Haar measure ensures that no two $w_a$ can coincide,
while the F-term constraint restricts the allowed pole structure to
the form a partition $\lambda^{(i)}$ of size
$|\lambda^{(i)}|=n_i$. Each partition $\lambda^{(i)}$ has an
associated 
Young tableau. We denote the coordinates of this Young tableau as
$(m,n)$ where 
$1\leq m \leq l(\lambda^{(i)})$ and $1\leq n\leq \lambda^{(i)}_m$  and
the set of poles are given by
\be \{\omega_a\}_i  = x_i\, q^{m+n-1}z^{m-n} \ \ \ {\rm with}\  (m,n)
\in 
\lambda^{(i)} \nn\ee
In this way, the integral \eqn{z} can be written as a sum over
coloured Young tableaux, coinciding with the usual expansion of the
Nekrasov partition function.  This argument works for $k<N$. When
$k>N$ there are further poles at the origin and more care must be
taken to massage the final answer into a sum over coloured
partitions. One way this can be done is to close the contour outside
instead by changing the integration variables from $w_a$ to
$\xi_a=1/w_a$ and observing that there is no pole at $\xi_a=0$. 

\subsection{A Single $U(N)$ Instanton}\label{nicesec}

The contour integration for the partition function can be carried out in terms of symmetric functions. However in most cases it is difficult to extract the information that we care about. 
One example where we can make progress is  the simple case of a single $U(N)$ instanton. Here we will derive an explicit expression for the partition function in terms of $SU(N)$
representations and show that it reproduces the results from canonical 
quantisation. More complicated cases will be discussed from a different viewpoint below.

\para
We will need some basic facts about Schur polynomials. For a given partition $\lambda=(\lambda_1,\lambda_2,\ldots,\lambda_N)$ with $\lambda_1\geq\lambda_2\geq\ldots\lambda_N>0$, the Schur polynomial $s_\lambda(X)$ in $N$ variables $X=(x_1,x_2,\ldots,x_N)$ is defined by
\be s_\lambda (X) = \sum_{\sigma\in S_N/S_N^\lambda}x_{\sigma(1)}^{\lambda_1}x_{\sigma(2)}^{\lambda_2}\ldots x_{\sigma (N)}^{\lambda_N}\prod_{i>j}\frac{1}{(1- {x_{\sigma (i)}}/{x_{\sigma (j)}})}\nn\ee
where $S_N^\lambda$ is the stabiliser of $\lambda$. 
Schur polynomials form a basis for the vector space of all symmetric functions, equipped with an orthonormal inner product
\be \left<s_\lambda,s_\mu\right>_S \equiv \frac{1}{N!}\left(\prod_{i=1}^N\frac{1}{2\pi i}\oint_C\right)\prod_{i\neq j}\left(1-\frac{x_i}{x_j}\right)s_\lambda (X)\,s_\mu (X^{-1}) = \delta_{\lambda,\mu} \nn\ee
The completeness of this basis is expressed through the Cauchy identity
\be \prod_{i=1}^n\prod_{j=1}^m\frac{1}{1-x_i y_j} = \sum_\lambda s_\lambda(X)\,s_\lambda(Y)\nn\ee
Starting from the contour integral of the partition function
\be {\cal Z}=
  \frac{1}{2\pi i}\oint\frac{dw}{w}\left[\frac{1-q^2}{(1-qz)(1-q/z)}\right]\frac{1}{w^k}\prod_{i=1}^N\frac{1}{(1-qw
    x_i)(1-q/w x_i)} \nn\ee
we note that $(1/2\pi i)\oint dw/w$ defines an orthonormal inner product on the
space of Schur polynomials in one variable $w$:
$\frac{1}{2\pi i}\oint\frac{dw}{w}s_\lambda (w)s_\mu(w^{-1}) =
\delta_{\lambda\mu}$. 
It is therefore natural to use
Cauchy's identity to rewrite the integral in terms of Schur
polynomials in $w$ and $X=(x_1,\ldots,x_N)$
\be
   \prod_{i=1}^N \frac{1}{1-qw x_i} &=& \sum_{p\geq 0}q^r s_{(p)}(w)\,
                                      s_{(p)}(X),\ \ \ \ \  w^{-k} = s_{(k)}(w^{-1}),\nn\\
  \prod_{i=1}^N \frac{1}{1-q/w x_i} &=& \sum_{r\geq 0} q^r
  s_{(r)}(w^{-1})\,s_{(r)}(X^{-1})\nn
  \ee
After using $s_{(r)}(w^{-1})s_{(k)}(w^{-1})=s_{(r+k)}(w^{-1})$ and
taking the inner product, which replaces $p$ with $r+k$, we get
\be {\cal Z} = \frac{(1-q^2)q^k}{(1-qz)(1-q/z)}\sum_{r\geq 0} q^{2r}
s_{(r+k)}(X)\,s_{(r)}(X^{-1})\nn\ee
This can be simplified further using the identity $s_{(r)}(X^{-1}) =
s_{(r^{N-1})}(X)/(\prod_i x_i)^r$ and the fact that $x_i$ are $SU(N)$
fugacities satisfying $\prod_i x_i =1$. Here, $(r^{N-1})$ denotes the partition of $(N-1)r$ with $r$ in all $N-1$ entries. The partition function now
reads
\be {\cal Z} = \frac{q^k (1-q^2)}{(1-qz)(1-q/z)}\sum_{r\geq 0}q^{2r}
s_{(r^{N-1})}(X)\,s_{(r+k)}(X) \nn\ee
After we multiply the factor $(1-q^2)$ into the sum, the coefficient
of the term $q^{k+2r}$ is
\be
s_{(r^{N-1})}(X)s_{(r+k)}(X)-s_{((r-1)^{N-1})}(X)s_{(k+r-1)}(X)= s_{(k+2r,r^{N-2})}(X)\nn\ee
where we use the Littlewood-Richardson rule to get the equality. In
physicists' language it goes as follows. Since Schur polynomials are characters of irreducible representations
of $SU(N)$, the above expression can be viewed in terms of Young
diagrams.
$s_{(r^{N-1})}(X)s_{(r+k)}(X)$ corresponds to
\be {\scriptstyle N-1}\left\{\begin{array}{c} \\ \\ \\ \end{array}\right. \!\!\!\!\overbrace{
{\yng(3,3,3,3)}}^r
\otimes\overbrace{\yng(5)}^{k+r}={\scriptstyle N-1}\left\{\begin{array}{c}\\ \\
                                                            \\ \end{array}\right. \!\!\!\!\overbrace{\yng(8,3,3,3)}^{k+2r}\oplus\left(\overbrace{\yng(2,2,2,2)}^{r-1}\otimes\overbrace{\yng(4)}^{k+r-1}\;\right)\nn\ee
where upon subtracting $s_{((r-1)^{N-1})}(X)s_{(k+r-1)}(X)$ which
corresponds to the second summand above, we are left
with the highest weight irreducible representation from the above
decomposition, whose character is $s_{(k+2r,r^{N-2})}(X)$ as claimed.
The final expression for the partition function is
\be {\cal Z} = \frac{q^k}{(1-qz)(1-q/z)}\sum_{r\geq 0} q^{2r} s_{(k+2r,r^{N-2})}(X) \label{oneinst-schur}\ee
Hence the ground state has energy $E_0 = k\omega$ and sits in the
${\rm Sym}^k({\bf N})$ representation of $SU(N)$. Moreover, we can
read off the whole spectrum from the partition function. For example,
excited states involving only $\varphi,\tilde{\varphi}$ excitations
have energy $E_r = \omega (k+2r)$, $r=1,2,3,\ldots$ sitting in the
$SU(N)$ representation
\be {\scriptstyle N-1}\left\{\begin{array}{c}\\ \\
                              \\ \end{array}\right.\!\!\!\!&\Yvcentermath1
                          \overbrace{\yng(8,3,3,3)}^{k+2r}\nn\ee
 In this way, we recover the spectrum we get from canonical quantisation in section \ref{nonabsec}.

\section{The Coulomb Branch Formula}\label{amisec}

The partition function \eqn{part} also goes by a different name: it is
the Hilbert series of the instanton moduli space. In other words, 
for $k=0$, it counts holomorphic functions on the instanton moduli
space graded by their charges under the global symmetries. For
$k\neq0$, it counts instead  holomorphic sections of the determinant
line bundle over the moduli space as described in \cite{yuji}.   
\para
A particularly powerful method  to compute Hilbert series 
invokes mirror symmetry of three-dimensional gauge theories
\cite{intseib}. 
This is a duality between two ${\cal N}=4$ supersymmetric gauge
theories, such that the Higgs branch of one theory is mapped to the
Coulomb branch of the other, and vice versa. The ADHM construction
naturally realises the moduli space of $n$ instantons as the Higgs branch
of an auxiliary $U(n)$ gauge theory. As the Higgs branch is dimension
independent the auxiliary theory can be taken to live in three
spacetime dimensions. Mirror symmetry then provides a dual realisation of the
same manifold as the Coulomb branch of another three-dimensional gauge
theory. It was shown in
\cite{ami1} that one can then recast the Hilbert series in terms of Coulomb branch
variables. Traditionally it is thought to be harder to work on the
Coulomb branch rather than 
the Higgs branch. For example, the former receives quantum
corrections, while the latter does not. 
However, to compute the Hilbert series, it turns out that one has
enough control over the relevant objects, specifically the monopole
operators, and one can use this to write a rather different expansion of the 
Hilbert series. 
\para
The Coulomb branch approach has been used to compute the Hilbert
series for many different Coulomb branches including, most pertinently, the instanton moduli spaces \cite{ami3}. 
The Coulomb branch approach to the partition function does not, at
present, allow for the inclusion of the effects of the five-dimensional 
Chern-Simons level $k$. We remedy this. First, however, 
we will review the Hilbert series for $n$ $U(N)$ instantons when $k=0$.

\subsection{$U(1)$ Instantons}

We start with $n$  instantons in $U(1)$. To this end, we consider a
three-dimensional $U(n)$ gauge theory, with an adjoint hypermultiplet
and a single fundamental hypermultiplet \cite{mirror}. The quiver is shown 
in the figure. The Higgs branch of this theory coincides with the
moduli space of $n$ $U(1)$ instantons, but this is something of a red
herring. Indeed, it turns out that this theory is self-mirror and, in
the limit that the gauge coupling $e^2\rightarrow \infty$, the Coulomb branch also
coincides with the moduli space of $n$ $U(1)$ instantons. It is this
Coulomb branch description that interests us. 

\para

 \EPSFIGURE{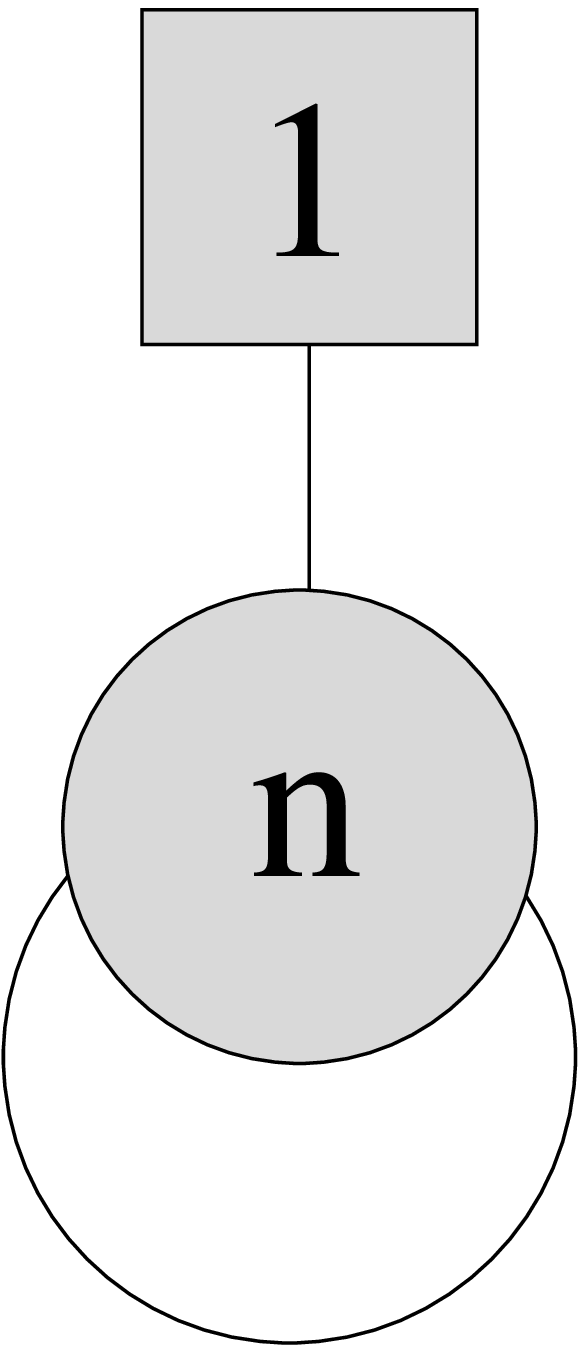,height=90pt}{}
\noindent
As we mentioned above, the  Coulomb branch suffers from quantum
corrections which often makes a detailed analysis difficult. Even the
simplest questions, such as 
the symmetries of the Coulomb branch, can be hard to answer. Theories
with ${\cal N}=4$ supersymmetry have an $SU(2)_R$ symmetry acting on
the Coulomb branch, one that is broken to $U(1)_R$ if we introduce masses for the
hypermultiplets. We identify this with the $U(1)_R$ of the instanton 
moduli space. 

\para
Other symmetries of the Coulomb branch arise from the shift of dual
photons. Instanton effects --- which are monopoles in three dimensions
---  break the shift symmetries of the dual photons arising from  $U(1)^{n-1}\subset
SU(n)$. This means that for each $U(n)$ factor of the gauge group, we get just a
single $U(1)$ global symmetry, with current given by $J^\mu =
\frac{1}{4\pi}\epsilon^{\mu\nu\rho}\,{\rm tr} f_{\mu\nu}$, where
$f_{\mu\nu}$ is the $U(n)$ field strength. This is sometimes referred to as a topological symmetry.
In the infra-red, it is not uncommon for these Abelian isometries to
be enhanced to non-Abelian symmetries of the Coulomb branch.
Indeed, this is expected to happen for the theory in the quiver, with the $U(1)$ topological
symmetry enhanced to $SU(2)_L$. 
\para
Our problem of computing the partition function \eqn{part} of our
matrix model has now turned into the problem of computing certain BPS
operators on the Coulomb branch of this three-dimensional gauge theory. 
Specifically, we are interested in operators that carry $SU(2)_L$
charge, which we have called $J$ above. From the Coulomb branch
perspective, these are {\it monopole operators}. 
 
\para
With this preamble in mind, here we present the prescription of
\cite{ami1,ami3} for counting monopole operators and, 
in doing so, computing the partition function of interest. For each
charge $J$, the monopole operators in $U(n)$ can sit in any of the
$U(1)^n$ Cartan elements.
To describe this decomposition, we introduce a sequence $\zeta$. 
\be \zeta: \zeta_1\geq \zeta_2\geq\ldots \zeta_n >-\infty\nn\ee
Note that $\zeta_a$ can be both positive or negative so $\zeta$ is
not, in general, a partition which includes only positive integers. 
The $\zeta_a$ are to be thought of as the charges of monopole
operators under $U(1)^n\subset U(n)$.
We require that 
\be |\zeta| := \sum_{a=1}^n \zeta_a = J\label{zetaj}\ee
We need two further definitions. We write
\be || \zeta || := \sum_{a=1}^n |\zeta_a|\nn\ee
We also define the {\it multiplicity} $m_l(\zeta)$ of $l$ in the sequence $\zeta$ to be the number of $\zeta_a$ that equal $l$, i.e.
\be m_l(\zeta) = \hash \{ a\, |\, \zeta_a=l\}\nn\ee
Let us write the partition function of $n$ instantons as ${\cal Z}_n(q,z)$. It can be expanded in  terms of $SU(2)_L$ quantum numbers as
\be {\cal Z}_n(q,z) = \sum_{J\in {\bf Z}} z^J\,Y_{n,J}(q)\label{zj}\ee
Note that $J$ is an angular momentum quantum number and takes both positive and negative values.  The result of \cite{ami1,ami3} for the partition function is 
 \be Y_{n,J}(q) = \sum_\zeta q^{||\zeta||}\prod_{l\in {\bf Z}}\prod_{a=1}^{m_l(\zeta)}\frac{1}{1-q^{2a}}\label{amifor1}\ee
 %

 \subsubsection*{Examples of $U(1)$ Instantons}
 
 It's useful to look at some examples. Let's start with a single $n=1$ instanton. In this case, the sequence of integers $\zeta$ is not much of a sequence: it consists of just a single integer which, by \eqn{zetaj}, is simply $J$ itself. The product over $l$ then has just a single term, coming from $l=J$. We have
 \be \mbox{1 instanton:}\ \ \ Y_{1,J}(q) = \frac{q^{|J|}}{1-q^2}
 \nn\ee
 From this we can compute the partition function  
\be {\cal Z}_1(q,z) = \sum_{J>0}\frac{1 + (qz)^J + (q/z)^J}{1-q^2}  = \frac{1}{1-qz}\frac{1}{1-q/z} \nn\ee
This coincides with the expected one-instanton partition function.

\para
For two instantons, the sequence is $\zeta_1\geq \zeta_2$ with $\zeta_1 + \zeta_2 = J$. We decompose the formula \eqn{amifor1} into two terms, one when $\zeta_1>\zeta_2$ and the other when $\zeta_1=\zeta_2$. We have
\be \mbox{2 instantons:}\ \ \ Y_{2,J}(q) = \sum_{\zeta_1>\zeta_2} \frac{q^{|\zeta_1|+|\zeta_2|}}{(1-q^2)^2} + \sum_{\zeta_1=\zeta_2}\frac{q^{2|\zeta_1|}}{(1-q^2)(1-q^4)}\nn\ee
We can then use \eqn{zj} to get the full partition function. 

\para
The above discussion was all for five dimensional 
Chern-Simons level $k=0$. We will discuss how one incorporates $k\neq
0$ in Section \ref{csaddsec} below.

\newpage

\subsection{$U(N)$ Instantons}

 \EPSFIGURE{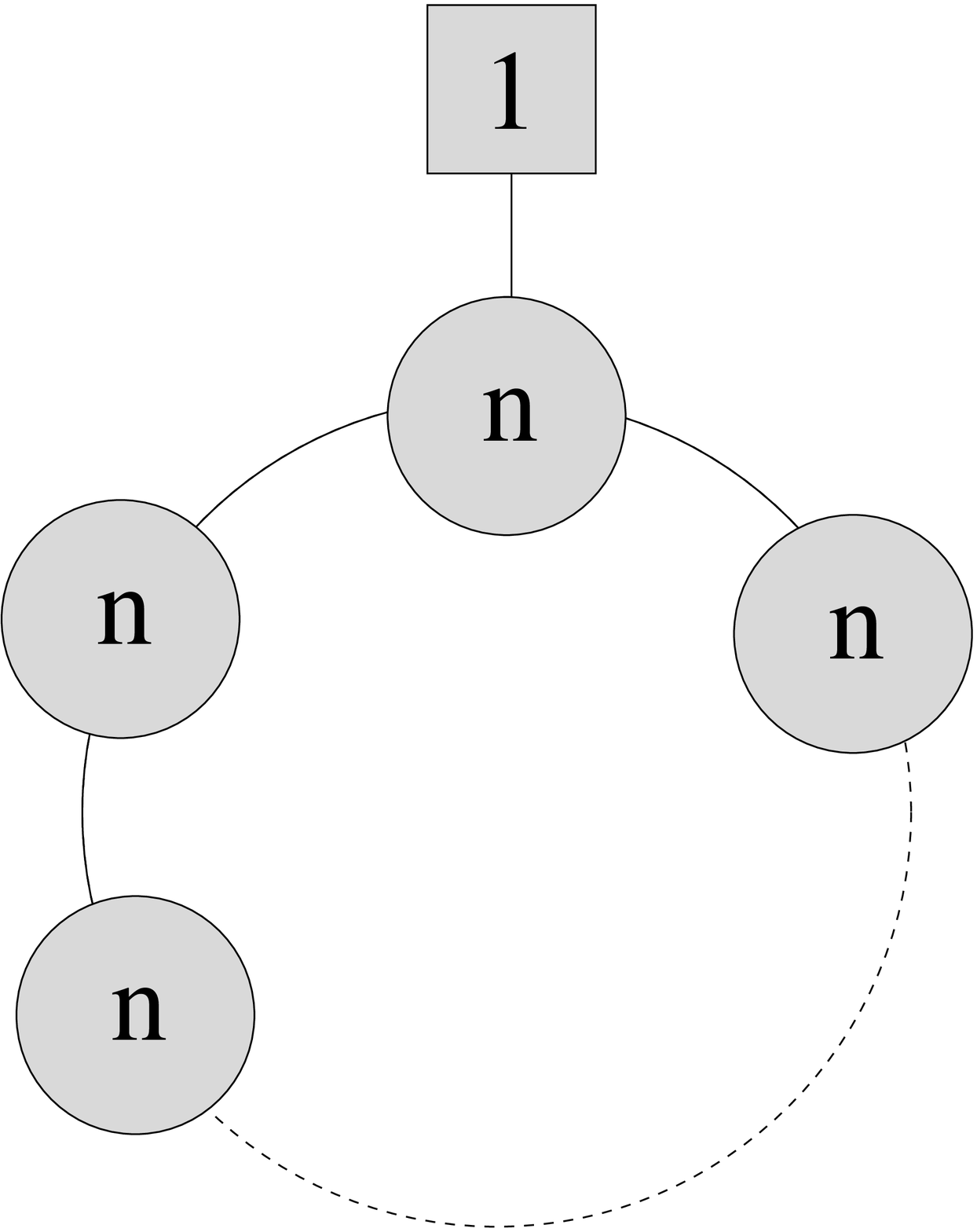,height=140pt}{}
\noindent
Let's now discuss the generalisation to $U(N)$ instantons. The
three-dimensional, ${\cal N}=4$ gauge theory whose Coulomb branch is
the moduli space of $n$ instantons in $U(N)$ is described by the
quiver shown on the right \cite{mirror}. 
It has gauge group $\prod_{i=1}^N U(n)_i$. There is a bi-fundamental hypermultiplet charged under consecutive gauge groups $U(n)_i\times U(n)_{i+1}$, where the periodicity of the quiver means that we identify $U(n)_{N+1} \equiv U(n)_1$. Finally, there is a single hypermultiplet that sits in the fundamental of $U(n)_N$. 

\para
As described previously, if the Coulomb branch has a non-Abelian 
symmetry in the infra-red, then only its Cartan subalgebra is visible
classically. For the quiver shown on the right, we have 
$N$ topological symmetries, one for each of the gauge group factors
$U(n)_i$. We denote the corresponding fugacities as $y_i$. Ultimately
these will be identified with the fugacities $z$ and $x_i$ for the
Cartan subalgebra of 
$SU(2)_L\times SU(N)$ which acts on the instanton moduli space; the 
map between them is
\be y_i = \frac{x_i}{x_{i+1}} \ \ \ i=1,\ldots, N-1\ \ \ \ 
{\rm and}\ \ \ \ y_N = z\frac{x_N}{x_{N-1}}\label{map}\nn\ee
This ensures that $\prod_{i=1}^N y_i = z$.

\para
To proceed, we again introduce sequences, $\zeta^i$, now one for each factor of the gauge group:
\be \zeta^i: \zeta^i_1\geq \zeta^i_2\geq\ldots \zeta^i_n >-\infty\nn\ee
As before, these describe the decomposition of monopole operators in $U(n)_i$. The formula of \cite{ami1,ami3} for the partition function is then:
\be {\cal Z}(q,y_i) = \sum_{\{\zeta^i\}}  q^{\Delta[\{\zeta^i\}]} \prod_{i=1}^N y_i^{|\zeta^i|} 
 \left[ \prod_{l\in {\bf Z}} \prod_{a=1}^{m_l(\zeta^i)} \frac{1}{1-q^{2a}}\right] \label{amiz}\ee
We can see that this is essentially a product of the contributions from monopole operators in each gauge group factor \eqn{amifor1}. These factors are tied together by the dimension of the monopole operator, which is given by
\be \Delta[\{\zeta^i\}] = -\frac{1}{2}\sum_{a,b=1}^n \sum_{i,j=1}^N  C_{ij} |\zeta_a^i-\zeta_b^j| + ||\zeta^N||\label{dim}\ee
Here $C_{ij}$ is the Cartan matrix of the affine Dynkin diagram that arises in the quiver which, in the present case, is 
\be C_{ij} = 2 \delta_{ij} - \delta_{i,j+1}  - \delta_{i+1,j}\nn\ee
where the periodic nature of the quiver again means that we identify $i\equiv i+N$. One can think of the first term as the contribution of vector bosons to the dimension of the monopole operator (they come with $-2$) and the second term as the contribution of the hypermultiplets (they come with +1). The final term in \eqn{dim} comes from the solitary hypermultiplet coupled to $U(n)_1$. 

\para
The Coulomb formula \eqn{amiz} also takes the form of an infinite sum. However, the term are now arranged in order of their energy, making it simple to read off the ground state energy, and the quantum numbers of the low lying states. 

\subsection{Adding the Chern-Simons Term}\label{csaddsec}

To describe the 4d quantum Hall states, we would like to generalise
the Coulomb branch formula \eqn{amiz} to include the five-dimensional 
Chern-Simons level $k$. 

\para

In the three-dimensional gauge theory whose Higgs
branch coincides with the instanton moduli space, this corresponds to
projecting onto operators of baryon number equal to $k$.\footnote{Note that this is not the same as introducing a Chern-Simons
  term in (either of) the three-
dimensional theories underlying the Higgs and Coulomb
  branch descriptions of the moduli space.}
In the mirror, Coulomb branch theory, one can mimic this by introducing $k$ units of magnetic flux for a suitable 
global symmetry, associated with  the periodic boundary condition around the circle in the standard brane construction of the affine
$\hat{A}_N$ quiver. For the  $\hat{A}_0$ quiver, this is just the symmetry which acts on the adjoint hypermultiplet.
The mirror deformation is then a particular shift of the topological charges as we go around the
periodic direction of the quiver. 

\para
This is rather straightforward to enact in the Coulomb branch formula \eqn{amiz}: the dimension of the monopole operator 
in \eqn{dim} is shifted to read
\be \Delta[\{\zeta^i\};k] = \sum_{i,j}\sum_{a,b}\left(-\delta_{ij}|\zeta^i_a-\zeta^j_b| + \delta_{i,j+1}|\zeta^i_a- \zeta^j_b+ k \delta_{j,N}|\right) + ||\zeta^N||\label{dim-cs}\ee
There are also some subtleties regarding the 
identification of the $SU(2)_L \times SU(N)$ and $U(1)_1\times\cdots\times U(1)_N$ fugacities. In terms of the former,
\be {\cal Z}(q,z,x_i) = \sum_{\{\zeta^i\}}  q^{\Delta[\{\zeta^i\};k]} z^{|\zeta^N|} \prod_{i=1}^N x_i^{|\zeta^i - \zeta^{i-1} + k\delta_{i,1}|} 
 \left[ \prod_{l\in {\bf Z}} \prod_{a=1}^{m_l(\zeta^i)} 
\frac{1}{1-q^{2a}}\right] \label{amiz-cs}\ee
Here we have chosen the shift to lie between the the nodes labelled
$N$ and $1$ and the formula is therefore not manifestly invariant under the
Weyl group of $U(N)$. Despite this the Weyl group is restored by the
sum over magnetic charges in our formula. 
Note that now the total weight in the fugacities $x_i$ is always $kn$,
in agreement with the canonical quantization picture. (Strictly, we are using
them here as $U(N)$ fugacities, and this measures the overall $U(1)$ charge 
enforced by the constraints in the theory.)
 \para
It is worth getting a feel for the effect of these modifications by 
exploring the physics encoded in \eqn{dim-cs}. In particular, 
we would like to make contact with the canonical quantization picture.

\subsubsection*{An Illuminating Example: Abelian Theories at $k=1$}

First, let us consider instantons in the simplest theory: $U(1)$ at level $k$. In Abelian theories, \eqn{dim-cs} simplifies significantly, leaving
\be \Delta[\zeta;k] = 2\sum_{a \neq b} \max\ \{0, |k|-|\zeta_a - \zeta_b|\} + |k|n + ||\zeta||\nn\ee
Clearly, the first term wants to space out the $\zeta_a$ with gaps of size $\ge|k|$, while the last term penalises placing them too far from $0$. In general, this creates an elaborate balancing act which gives rise to the subtle pattern of $SU(2)$ irreps that we saw in Section \ref{abeliansec}.

\para
We will satisfy ourselves with observing what happens for $k=1$. Here, the only relevant contributions to the energy are  a cost of $+2$ for each pair of $\zeta_a$ which are the same, and an overall cost of $\sum_a|\zeta_a|$. But this precisely parallels the canonical quantization picture, with the interpretation that a particular $\zeta$ corresponds to a baryon that looks like
\be
\zeta \ \  \longleftrightarrow \ \  \left|\psi\right> \sim \left[ \prod_{l \ge 0} \prod_{a=1}^{m_l(\zeta)} Z^{l+a}\tilde{Z}^{a} \varphi \right]^\dagger\left[ \prod_{l < 0} \prod_{a=1}^{m_l(\zeta)} Z^{a}\tilde{Z}^{l+a} \varphi \right]^\dagger \left| 0\right>
\nn\ee
Notice that indeed the energy of this baryon is precisely $\Delta[\zeta;k]$, and also that the total $U(1) \subset SU(2)$  charge is $\sum_l l m_l(\zeta) = |\zeta|$.

\para
The only remaining terms in \eqn{amiz-cs} to understand are the tower of excited states of the same $SU(2)$ charge encoded in the final bracket. But these simply encapsulate the possibility of adding extra $Z\tilde{Z}$ factors into each term of the baryon, correctly taking account of the combinatorics when there are terms in the baryon which differ by a power of $Z\tilde{Z}$.

\para
It is slightly subtle to understand larger values of $k$ because $\zeta$ measures quantities which are summed over all baryons, so one must work out how to partition things up between baryons. However,  the basic idea described above carries over, with $\zeta$ encoding the $SU(2)$ quantum numbers of the terms in the baryons.
One can also extend this to the non-Abelian case; the $x_i$ fugacities are a guide to decide which $\varphi_i, \tilde{\varphi}_j$ to insert for the initial state analogous to $\left|\psi\right>$ above. Then one enumerates not only $Z\tilde{Z}$ powers as excitations, but also $\varphi_i \tilde{\varphi}_i$ powers.

\para
In this way, the  Coulomb branch parametrization of states in terms of the sequences $\zeta$ parallels  the description of baryons seen in  canonical quantization.  We have essentially gone full circle, from canonical quantization to a Molien-type integral, to a Higgs branch partition function, to a Coulomb branch partition function, and back to the canonical picture once more.

\subsubsection*{Another Example: Schur Polynomials for 1 instanton in $U(2)$}

We can also try to make contact with the expression \eqn{oneinst-schur} for the one instanton partition function, as decomposed into its irrep constituents. This is non-trivial because the Coulomb branch expression does not have a very transparent relationship with the $U(N)$ symmetry of the original theory.

\para
Because of this, it is most practical to restrict ourselves to the case $N=2$. The Coulomb branch formula then reads
\be {\cal Z} = \frac{1}{(1-q^{2})^2} \sum_{\zeta^1,\zeta^2}  q^{|\zeta^2-\zeta^1| + |\zeta^1-\zeta^2 + k| + |\zeta^2|} z^{\zeta^2} \prod_{i=1}^N x_i^{\zeta^i - \zeta^{i-1} + k\delta_{i,1}}  \nn\ee
We can now write $s = \zeta^1 - \zeta^2$ and do the $\zeta^2$ sum. This reduces the expression to 
\be {\cal Z} = \frac{1}{(1-q^{2})(1-qz)(1-q/z)} \sum_{s}  q^{|s| + |s + k|} \left(\frac{x_1}{x_2}\right)^s x_1^{k}  \nn\ee
To get to the final result, we split the $s$ summation up into the ranges $s \ge 0$, $s < -k$, and what remains. One may easily verify that combining these terms together and multiplying by the series expansion for $(1-q^2)$ gives
\be {\cal Z} &=& \frac{1}{(1-qz)(1-q/z)} \sum_{s=0}^\infty  q^{2s+k} x_1^{k} \left[ \left(\frac{x_1}{x_2} \right)^{s} + \left(\frac{x_1}{x_2} \right)^{s-1} + \cdots + \left(\frac{x_1}{x_2} \right)^{-s-k} \right] \nn\\
 &=& \frac{1}{(1-qz)(1-q/z)} (x_1 x_2)^{k/2} \sum_{s=0}^\infty  q^{2s+k} \frac{s_{(2s+k)}(x_1, x_2)}{(x_1x_2)^{(2s+k)/2}} \nn\ee
with the correct total $U(1)$ charge; alternatively, setting $x_1x_2=1$ we get
\be{\cal Z} = \frac{q^k}{(1-qz)(1-q/z)} \sum_{s=0}^\infty  q^{2s}s_{(2s+k)}(x_1, x_2) \ee
in precise agreement with \eqn{oneinst-schur}.

\para
Thus the Schur form, which makes explicit the $SU(N)$ structure of the partition function, is obtained from the Coulomb branch expression only upon doing various non-trivial manipulations. It would be nice to understand the $SU(N)$ structure of the spectrum more directly from the Coulomb branch formula.

\subsection{Derivation of the Coulomb Branch Formula}

In this section we derive  the  Coulomb branch formula \eqn{amiz} and \eqn{dim-cs} from the original  Molien integral form of the partition function \eqn{z} coincide. To proceed, we will need a little bit of symmetric functionology; the canonical reference for this material is \cite{Macdonald:95}.

\subsubsection*{Some Symmetric Function Notation}

To start,  we define the set of variables $W=\{w_1,\dots, w_n\}$. Associated to each partition $\lambda$ with the length  $\ell(\lambda)\leqslant n$ there is a non-zero symmetric polynomial in $W$ over the field $\mathbb Q(t)$ (finite Laurent polynomials in $t$ with rational coefficients) called the Hall-Littlewood polynomial, 
\be
P_\lambda(W;t):=\frac{1}{N_\lambda(t)}\sum_{\sigma\in S_n}\sigma\left(\prod_lw_l^{\lambda_l}\prod_{l<m}\frac{w_l-tw_m}{w_l-w_m}\right)\,,\nonumber
\ee
where $S_n$ is the symmetry group acting on the indices of the $W$ variables by permutations. Here $N_\lambda$ is a normalisation factor defined by
\be
N_\lambda(t):=\frac{\varphi_{n-\ell(\lambda)}(t)\prod_{j\geqslant 1}\varphi_{m_j(\lambda)}(t)}{(1-t)^n}\,,\nonumber
\ee
where
\be \varphi_a(t)=\prod_{j=1}^a(1-t^j)\  \ \ {\rm and}\ \ \  m_j(\lambda)=|\{i\geqslant 1:\lambda_i=j\}| \nonumber\ee
The Hall-Littlewood polynomials with variables $W$ and $t$ form a basis for all symmetric functions in $W$ over the field $\mathbb Q(t)$. We have an inner product on symmetric polynomials defined by the integral
\be\label{DotProd}
\left< P_\lambda,P_\mu\right>_t&:=&\oint_Cd\mu[W;t]\ P_\lambda(W;t)\,P_\mu(W^{-1};t) \\
&:=&\frac{1}{n!}\left(\prod_{l=1}^n\frac{1}{2\pi i}\oint_C\frac{dw_l}{w_l}\right)\prod_{l\neq m}\frac{1-w_l/w_m}{1-tw_l/w_m}\ P_\lambda(W;t)\,P_\mu(W^{-1};t)\nonumber\\
&=&\frac1{N_\lambda(t)}\delta_{\lambda\mu}\nonumber\,.
\ee
We can further generalise our polynomials to be a basis for all finite symmetric Laurent polynomials in $W$ over $\mathbb Q(t)$. To do this we first define the set of $n$ ordered integers 
\be\label{defM}
\mathfrak M:=\{(\zeta_1,\dots,\zeta_n)\in\mathbb Z^n|\zeta_1\geqslant \zeta_2\geqslant\dots\geqslant \zeta_n\}\cong \mathbb Z^n/S_n\,.
\ee
For any $\zeta\in \mathfrak M$ we can define a partition $\lambda$ via $\lambda_l:=\zeta_l-\zeta_n$ for $l=1,\dots,n-1$ and $\lambda_n=0$, then we can define
\be
N_\zeta(t):=N_\lambda(t)\,,\nonumber
\ee
and we define the shifted Hall-Littlewood polynomials by
\be\label{shiftedHL}
\Psi_\zeta(W;q):=\prod_lw_l^{\zeta_n}\,P_\lambda(W;q)\,.
\ee
One can quite easily show that given these definitions we have, using the same inner product,
\be\label{innprod}
\left<\Psi_\zeta,\Psi_\eta\right>_t=\frac{1}{N_\zeta(t)}\delta_{\zeta\eta}\,.
\ee
A simple trick, that we will use extensively, is the following: for $\zeta,\eta\in \mathfrak M$, we can write a product of two of these polynomials shifted by an arbitrary constant $a\in\mathbb Z$, namely
\be\label{partTrick}
\Psi_\zeta(W)\Psi_\eta(W^{-1})=\Psi_{\zeta+(a^n)}(W)\Psi_{\eta+(a^n)}( W^{-1})
\ee
We have the Cauchy identity
\be\label{Cauchy}
\prod_{l,m}\frac{1-tx_ly_m}{1-x_ly_m}=\sum_{\lambda}P_\lambda(X;t)Q_\lambda(Y;t)\,.
\ee

\subsubsection{Back to the Partition Function}

The machinery above can be brought to bear on our partition function. Our original expression for the partition function \eqn{z} is
\be
\mathcal Z(k)&=&\frac{1}{n!}\prod_{l=1}^n\left(\oint\frac{dw_l}{2\pi iw_l^{k+1} }\prod_{i=1}^N\frac{1}{(1-qw_lx_i)(1-q/w_lx_i)}\right)\nonumber\\
&&\ \ \ \ \times \prod_{l\neq m}(1-w_l/w_m)\prod_{l,m=1}^n\frac{1-q^2w_l/w_m}{(1- qw_l/zw_m)(1-qzw_l/w_m)}\nonumber\,.
\ee
Here we have explicitly written the dependence on the background baryonic charge $k$.

\para
To start, we write the second line as 
\be
\prod_{l\neq m}(1-w_l/w_m)&& \!\!\!\prod_{l,m=1}^n\frac{1-q^2w_l/w_m}{(1-\frac qzw_l/w_m)(1-qzw_l/w_m)}\nonumber\\
=&&\frac{1}{(1-q^2)^n}\prod_{l\neq m}\frac{1-w_l/w_m}{1-q^2w_l/w_m}\prod_{l,m}\frac{(1-q^2w_l/w_m)(1-q^2w_l/w_m)}{(1-qzw_l/w_m)(1-\frac qzw_l/w_m)}\nonumber\,,
\ee
and define functions 
\be
Q[W,\tilde W;z,q]&:=&\prod_{l,m}\frac{(1-q^2w_l/w_m)(1-q^2\tilde w_l/\tilde w_m)}{(1-qzw_l/\tilde w_m)(1-\frac{q}{z}\tilde w_l/w_m)}\nn\ee
and
\be
\pi_f(W;x)&:=&\prod_{l=1}^N\frac{1}{1-qw_lx},\quad \pi_{\bar f}(W;x):=\prod_{l=1}^N\frac{1}{1-q/w_lx}\,.\nonumber
\ee
The partition function can then be written as
\be
\mathcal Z(k)=\frac{1}{(1-q^2)^n}\oint_Cd\mu[W;q^2]\prod_{l=1}^nw_l^{-k}\left(\prod_{i=1}^N\pi_f(W;x_i,q)\pi_{\bar f}(W;x_i,q)\right)Q[W,W;z,q]\nonumber\,.
\ee
Here the contour $C$  is defined to be the unit $n$-torus.

\para
To convert our Higgs branch expression into a Coulomb branch expression for the mirror dual theory, we will make  use of ``Dirac delta functions" for symmetric polynomials. These allow us to rewrite the Hilbert series  as an integral over the maximal torus of the gauge group of the affine quiver, $T(U(n)^{N})\cong T^{nN}\cong C^N$. 
To do this we first define the function
\be
K[W,\tilde W]:=\sum_{\zeta\in \mathfrak M}N_\zeta(q^2)\Psi_\zeta(W;q^2)\Psi_\zeta(\tilde W^{-1};q^2)\,,\nonumber
\ee
where  $\mathfrak M$ is the set of $n$ ordered integers defined in equation (\ref{defM}) and the shifted Hall-Littlewood polynomials $\Psi$ are defined in equation (\ref {shiftedHL}). Then, using the meaure \eqn{DotProd}, we have for any symmetric Laurent polynomial $f$ in $W$ over $\mathbb Q(q^2)$ 
\be
f(W)=\oint_Cd\mu[\tilde W;q^2]\ K[W,\tilde W]\,f(\tilde W)\,.\nonumber
\ee
This follows from the fact that the functions $\Psi_\zeta$ form a linear $\mathbb Q(q^2)$-basis for symmetric finite Laurent polynomials and the orthogonality property outlined in equation (\ref{innprod}). There is a notational subtlety here;  the sum should not be inside the integral, but rather
\be
f(W)=\sum_{\zeta\in \mathfrak M}\oint_Cd\mu[\tilde W;q^2]\ N_\zeta(q^2)\,\Psi_\zeta(W;q^2)\,\Psi_\zeta(\tilde W^{-1};q^2)\,f(\tilde W)\,.\nonumber
\ee
It is understood that whenever we write $K$ we mean that the sum sits outside the integral. This is akin to how the Dirac delta function is not well defined as a function unless inside an integral.

\para
Defining $W^{(0)}\equiv W$ and $W^{(N+1)}\equiv\tilde W$, we insert a complete set of states to rewrite the integral as 
\be
\mathcal Z(k)&=&\frac{1}{(1-q^2)^n}\oint_Cd\mu[W;q^2]\oint_Cd\mu[\tilde W;q^2]\ \prod_{i=1}^N\oint_Cd\mu[W^{(i)};q^2]\nonumber\\
&&\ \ \ \ \ Q[W,\tilde W]\prod_{i=1}^N\pi_f(W^{(i)},x_i)\pi_{\bar f}(W^{(i)},x_i)\prod_{i=1}^{N+1}K[W^{(i-1)-1},W^{(i)-1}]\nonumber
\ee
This motivates us to define, for $\zeta,\eta\in \mathfrak M$,
\be
M_{\zeta\eta}(x):=N_\zeta(q^2)\oint_Cd\mu[W;q^2]\ \Psi_\zeta(W;q^2)\,\pi_f(W;x,q)\,\pi_{\bar f}(W;x,q)\,\Psi_\eta(W^{-1};q^2)\,,\label{originalM}\nn\ee
and
\be
\mathcal O_{\zeta\eta}(k):=\frac{N_\zeta(q^2)}{(1-q^2)^n}\oint_Cd\mu[W;q^2]\oint_Cd\mu[\tilde W;q^2]\ \Psi_{\zeta-(k^n)}(W)\,\Psi_\eta(\tilde W^{-1})\,Q[\tilde W,W]\,,\nonumber
\ee
and the trace is over the vector space of finite symmetric Laurent polynomials, which is spanned by $\{\Psi_\zeta(W;q^2)|\zeta\in \mathfrak M\}$. Note once again that the sums over $\mathfrak M$ are outside the integrals. The partition function can then be written in the simple form
\be
\mathcal Z(k)=\tr_{\mathcal H}\,\big[\mathcal O(k)M(x_1)M(x_2)\dots M(x_N)\big]\,.\nonumber
\ee

\para
The problem has been reduced to evaluating $M_{\zeta\eta}(x)$ and $\mathcal O_{\zeta\eta}(k)$. We start with $M_{\zeta\eta}(x)$ and use the trick in equation (\ref{partTrick}) to restrict to the case where all the elements of $\zeta$ and $\eta$ are negative so that we can write 
\be 
M_{\zeta\eta}(x) &=&N_\zeta(q^2)\oint_{C}d\mu[W;q^2]\ \Psi_\zeta(W;q^2)\,\pi_{\bar f}(W)\,\sum_{\chi\in\mathcal P_n}\oint_{C}d\mu[\tilde W;q^2]\nonumber\\
&&\ \ \ \ \  \ \ \ \ \ \frac1{(1-q^2)^n}\,P_\chi(W;q^2)\,Q_\chi(\tilde W^{-1};q^2)\,\Psi_\eta(\tilde W^{-1};q^2)\pi_f(\tilde W)\,.\nonumber
\ee
The set $\mathcal P_n$ is the set of all partitions of length $\leqslant n$. 

\para
Next, we swap the sum with the integral, use the Cauchy identity (\ref{Cauchy}) and expand $\Psi_\zeta$ and $\Psi_\eta$ as a sum over permutations so that the integrand in the summand is a rational function. This gives
\be\label{2int}
M_{\zeta\eta}(x) = \frac{1}{N_\eta(q^2)(1-q^2)^n}\prod_{l=1}^n && \!\left(\oint_C\frac{d w_l}{2\pi i w_l}\oint_{\tilde C}\frac{d\tilde w_l}{2\pi i\tilde w_l} \frac{w_l^{\zeta_l}}{1-q/w_lx}\frac{\tilde w_l^{-\eta_l}}{1-q\tilde w_lx}\right) \\
&&\!\!\!\! \times \prod_{l,m=1}^n\frac{1-q^2 w_l/\tilde w_m}{1-w_l/\tilde w_m}    \prod_{l<m}\left(\frac{1-\tilde w_m/\tilde w_l}{1-q^2\tilde w_m/\tilde w_l}\frac{1- w_l/ w_m}{1-q^2 w_l/w_m}\right)\,.\nonumber
\ee
An important subtlety to note here is that the contour for $\tilde W$ has changed to $\tilde C:=(1+\varepsilon)C$ for some sufficiently small   $\varepsilon>0$, so that it doesn't interfere with the rest of the pole structure. This is because the denominator, $1-w_l/\tilde w_m$, that arises from the Cauchy identity after swapping the sum of partitions with the integral over $\tilde W$ means that the form we integrate is not defined on any contour for $w_l$ and $\tilde w_m$ such that these contours intersect. The justification for this choice of contour comes from noting that, if we evaluate the poles for $\tilde w_l$ in the order $\tilde w_n, \tilde w_{n-1},\dots,\tilde w_1$, and choose poles within the contour, then  we find that the only poles  are parameterised by a permutation $\sigma\in S_n$ with $\tilde w_l=w_{\sigma(l)}$. (Recall that  $\eta_l<0$ so there are no poles at $0$.) Upon evaluating the residue we then find exactly the original expression for $M_{\zeta\eta}(x)$.
\para

Considering the relative ordering of the decreasing sequences $\zeta,\eta\in \mathfrak M$, we redefine the labels as follows
\be
\zeta_l\equiv \zeta_{i,\alpha_i}&\quad \mbox{ for }i=1,\dots,i_{\tiny{\mbox{max}}},\;\alpha_i=1,\dots,n_i\,,\nonumber\\
\eta_l\equiv \eta_{i,\tilde \alpha_i}&\quad \mbox{ for }i=1,\dots,i_{\tiny{\mbox{max}}},\;\tilde\alpha_i=1,\dots,\tilde n_i\,,\nonumber
\ee
such that $(n_1,\dots,n_{i_{\tiny{\mbox{max}}}})$ and $(\tilde n_1,\dots,\tilde n_{i_{\tiny{\mbox{max}}}})$ are compositions of $n$ (with $n_1$ and $\tilde n_{i_{\tiny{\mbox{max}}}}$ possibly zero, but all other values strictly positive integers) and
$\zeta_{i,\alpha_i}\leqslant \zeta_{i,\beta_i}$ $\forall\; \alpha_i\geqslant \beta_i$, and $\eta_{i,\tilde\alpha_i}\leqslant \eta_{i,\tilde\beta_i}$ $\forall \;\tilde\alpha_i\geqslant \tilde\beta_i$ and, finally, $\zeta_{i,\alpha}\geqslant \eta_{i,\tilde\alpha}\geqslant \zeta_{i+1,\alpha'}$ $\forall\,\alpha,\tilde\alpha,\alpha'$.
%
%
%
With these definitions, we can now evaluate the matrix elements. We need the following lemmas:

\para
{\bf Lemma 1:} The value of $M_{\zeta\eta}(x)$ is saturated by the pole located at
\be
w_l=\frac{q^{2s_l}}{x}\ \ \ {\rm and} \ \ \ \tilde w_l=\frac{q^{2\tilde s_l}}{x}\,,\nonumber
\ee
where
\be
s_l&=&s_{i,\alpha}=\sum_{j=1}^{i-1}(n_j-\tilde n_j)+\alpha-\frac{1}{2}\,,\nonumber\\
\tilde s_l&=&\tilde s_{i,\tilde\alpha}=\sum_{j=1}^in_j-\sum_{j=1}^{i-1}\tilde n_j+\frac{1}{2}-\tilde\alpha\,.\nonumber
\ee
So if $\zeta_1\geqslant \eta_1,\; s_1=\frac{1}{2}$ and if $\eta_1>\zeta_1, \;\tilde s_1=-\frac{1}{2}$.
Furthermore the residue of this pole is such that we have
\be
M_{\zeta\eta}(x)=\frac{1}{\prod_{a\in\mathbb Z}\varphi_{m_a(\eta)}(q^2)}x^{|\eta|-|\zeta|}q^{\Delta_1[\zeta,\eta]}\,,\nonumber
\ee
where $\Delta_1[\zeta,\eta]$ is defined in equation (\ref{delta1}). This is the topic of Appendix \ref{Mmn}.

\para
Note that this expression for $M_{\zeta\eta}(x)$ is invariant under shifts of the form $(\zeta,\eta)\mapsto(\zeta+(c^n),\eta+(c^n))$ for all $c\in\mathbb Z$. This is what we would expect as we could see this symmetry already when we used equation (\ref{partTrick}).

\para
For the other term $\mathcal O_{\zeta\eta}(k)$, we have that

\para
{\bf Lemma 2:} 
\be
\mathcal O_{\zeta\eta}(k)=\delta_{\zeta-(k^n)\,\eta}z^{|\eta|}q^{||\eta||}\nonumber
\ee
where $||\eta||:=\sum_{l=1}^n|\eta_l|$ and $|\eta|:=\sum_{l=1}^n\eta_l$. We prove this in Appendix \ref{Omn}.

\para
Given these lemmas we can write the Hilbert series as
\be
\mathcal Z(k)=\sum_{\vec \zeta\in \mathfrak M^N}z^{|\zeta^{(1)}|}q^{||\zeta^{(1)}||}M_{\zeta^{(1)}\zeta^{(2)}}(x_1)M_{\zeta^{(2)}\zeta^{(3)}}(x_2)\dots M_{\zeta^{(N)}\zeta^{(1)}+(k^n)}(x_N)\,.\nonumber
\ee
To see that this does indeed agree with (\ref{amiz-cs}) we shall expand out the $M's$. First we consider the bare dimension, written in equation (\ref{dim-cs}). Our expression for the bare dimension is (defining $\zeta^{(N+1)}\equiv\zeta^{(1)}$)
\newpage
\be
\Delta[\{\zeta\};k]&=&\sum_{i=1}^N\Delta_1[\zeta^{(i)},\zeta^{(i+1)}+\delta_{i\,N}(k^n)]+||\zeta^{(1)}||\nonumber\\
&=&\sum_{i=1}^N\sum_{l,m=1}^n\left(|\zeta^{(i)}_l-\zeta^{(i+1)}_m+\delta_{iN}k|-\frac12|\zeta^{(i)}_l-\zeta^{(i)}_m|-\frac12|\zeta^{(i+1)}_l-\zeta^{(i+1)}_m|\right)+||\zeta^{(1)}||\nonumber\\
&=&\sum_{i=1}^N\sum_{l,m=1}^n\left(|\zeta^{(i)}_l-\zeta^{(i+1)}_m+\delta_{iN}k|-|\zeta^{(i)}_l-\zeta^{(i)}_m|\right)+||\zeta^{(1)}||\nonumber
\ee
We then redefine our variables via $\zeta^{(i)}=\zeta^{N+1-i}$ and  write
\be
\Delta[\{\zeta\};k]&=&\sum_{i=1}^N\sum_{l,m=1}^n\left(|\zeta^i_l-\zeta^{i-1}_m+\delta_{i1}k|-|\zeta^{i}_l-\zeta^{i}_m|\right)+||\zeta^{N}||\nonumber\\
&=&\sum_{i=1}^N\sum_{l,m=1}^n\left(|\zeta^{i}_l-\zeta^{i-1}_m+\delta_{i1}k|-|\zeta^{i}_l-\zeta^{i}_m|\right)+||\zeta^{N}||\nonumber\\
&=&\sum_{i,j=1}^N\sum_{l,m=1}^n\left(\delta_{i\,j+1}|\zeta^{i}_l-\zeta^{j}_m+\delta_{jN}k|-\delta_{i\,j}|\zeta^{i}_l-\zeta^{j}_m|\right)+||\zeta^{N}||\nonumber
\ee
This exactly the expression in equation (\ref{dim-cs}). We have
\be
\mathcal Z(k)=\sum_{\vec \zeta\in \mathfrak M^N}z^{|\zeta^{N}|}q^{\Delta[\{\zeta\};k]}\prod_{i=1}^N x_i^{|\zeta^{N-i+1}-\zeta^{N-i}+k\delta_{iN}|}\prod_{i=1}^N\prod_{a\in\mathbb Z}\frac1{\varphi_{m_a(\zeta^i)}(q^2)}\,.\nonumber
\ee
Upon acting with an element of the Weyl group of $U(N)$ such that $x_i\mapsto x_{N-i+1}$ we find we exactly reproduce equation (\ref{amiz-cs}),
\be
\mathcal Z(k)=\sum_{\vec \zeta\in \mathfrak M^N}z^{|\zeta^{N}|}q^{\Delta[\{\zeta\};k]}\prod_{i=1}^N x_i^{|\zeta^{i}-\zeta^{i-1}+k\delta_{i1}|}\prod_{i=1}^N\prod_{a\in\mathbb Z}\frac1{\varphi_{m_a(\zeta^i)}(q^2)}\,.\nonumber
\ee

\newpage

\appendix

\section{Appendix: Classical Matrix Model Solutions}

It is possible to see the structure of the baryons in the quantum theory, presented in \eqn{lowestb}, from some solutions to the classical matrix model. Recall these baryons look like this:
\be B  = \epsilon^{a_1\ldots a_n} \varphi_{a_1} (Z\vp)_{a_2}(\tz\vp)_{a_3} (Z^2\vp)_{a_4}\ldots \ldots(Z^{l_n}\tz^{m_n}\varphi)_{a_n}\nn\ee

\para
A simple set of operators which carry information about this state is given by
\be O_{mn} = \frac{1}{m!n!} {\rm Tr} \ Z^{\dagger m} \tilde{Z}^{\dagger n} Z^m \tilde{Z}^n \ee
Essentially, because of the minimal nature of the ground state baryon, $O_{mn}$ counts only the number of ways of choosing $m$ $Z$ terms and $n$ $\tilde{Z}$ operators from a {\it single term} in the above baryon.

\para
Using this, one can now calculate $\hat{O}_{mn}$, the number of terms in the baryon which are {\it exactly} $Z^m \tilde{Z}^n \varphi$. (Clearly, for a single baryon, this could be taken to be either $0$ or $1$.) One does this by subtracting off the contributions from terms with $m'>m$ or $n'>n$ which also contribute to $O_{mn}$. This is easy to work out; the $Z^{m+i} \tilde{Z}^{n+j} \varphi$ terms contribute an extra ${m+i \choose m }{n+j \choose n} \hat{O}_{m+i \ n+j}$ to $O_{mn}$. Therefore, one knows that the largest $(m,n)$ (with respect to the obvious partial ordering) with a non-vanishing $O_{mn}$ is correct and so $\hat{O}_{mn} = O_{mn}$, and then one can calculate the next-largest $\hat{O}$ using the above observation, and so on.

\para
Now translate this to the classical ground state. For example, for $n=3$ particles, we have (setting $k=1$)
\be
Z = \left(\begin{array}{ccc} \ 0 \ & \ 0 \ & \ 0 \ \\ 0 & 0 & 1 \\ 0 & 0 & 0  \end{array}\right)
\quad {\rm and} \quad
W = \left(\begin{array}{ccc} \ 0 \ & \ 0 \ & \ 1 \ \\ 0 & 0 & 0 \\ 0 & 0 & 0  \end{array}\right)
\quad {\rm with } \quad
\varphi = \left(\begin{array}{c} 0 \\  0\\ \sqrt{3}\end{array}\right)
\nn\ee
setting $\tilde{\varphi} = 0$. One immediately finds that
\be
O_{mn} = \left(\begin{array}{cc} \ 3 \ & \ 1 \ \\ 1 & 0 \end{array}\right)
\nn\ee
and so, subtracting off the excess contributions,
\be
\hat{O}_{mn} = \left(\begin{array}{cc} \ 1 \ & \ 1 \ \\ 1 & 0 \end{array}\right)
\nn\ee
This indeed matches with the ``$\varphi Z\varphi \tilde{Z} \varphi$'' structure of the quantum ground state.

\para
Moreover, if one applies a simple $SU(2)$ rotation to $(Z,W)$, one finds that $\hat{O}_{mn}$ is invariant, as one would expect, since the quantum state is an $SU(2)$ singlet.

\para
Let us now choose a more interesting state, at $n=5$. A classical, minimal energy configuration is now
\be
Z = \left(\begin{array}{ccccc} \ 0 \ &&&& \\ & \ 0 \ &&1& \\ && \ 0 \ && \\ &&& \ 0 \ &\sqrt{2} \\ &&&& \ 0 \   \end{array}\right)
\quad {\rm and} \quad
W = \left(\begin{array}{ccccc} \ 0 \ &&1&& \\ & \ 0 \ &&& \\ && \ 0 \ &&\sqrt{2} \\ &&& \ 0 \ & \\ &&&& \ 0 \   \end{array}\right)
\quad {\rm with } \quad
\varphi = \left(\begin{array}{c} 0 \\ 0 \\0 \\  0\\ \sqrt{5}\end{array}\right)
\nn\ee

\para
Now we find
\be
O_{mn} = \left(\begin{array}{ccc} \ 5 \ & \ 3 \ & \ 1 \ \\ 3 & 0 & 0 \\ 1 &0&0 \end{array}\right)
\quad\implies\quad
\hat{O}_{mn} = \left(\begin{array}{ccc} \ 1 \ & \ 1 \ & \ 1 \ \\ 1 & 0 & 0 \\ 1 &0&0 \end{array}\right)
\nn\ee
This corresponds to the  ``$\varphi Z\varphi \tilde{Z} \varphi Z^2\varphi \tilde{Z}^2 \varphi$'' choice of quantum state.

\para
However, this is no longer invariant under the action of $SU(2)$. Indeed, acting with a rotation by $\theta$ from the natural $SO(2) \subset SU(2)$ upon $Z,W$ leads to the solution
\be
Z = \left(\begin{array}{ccccc} \ 0 \ &&-\sin\theta&& \\ & \ 0 \ &&\cos\theta& \\ && \ 0 \ &&-\sqrt{2} \sin \theta \\ &&& \ 0 \ &\sqrt{2} \cos\theta \\ &&&& \ 0 \   \end{array}\right)
\quad {\rm and} \quad
W = \left(\begin{array}{ccccc} \ 0 \ &&\cos\theta&& \\ & \ 0 \ &&\sin\theta& \\ && \ 0 \ &&\sqrt{2} \cos \theta \\ &&& \ 0 \ &\sqrt{2} \sin\theta \\ &&&& \ 0 \   \end{array}\right)
\nn\ee
with a corresponding 
\be
\hat{O}_{mn} = \left(\begin{array}{ccc} \ 1 \ & \ 1 \ & \ \frac{1+\cos^2 2\theta}{2} \ \\ 1 & \sin^2 2\theta & 0 \\ \frac{1+\cos^2 2\theta}{2} &0&0 \end{array}\right)
\nn\ee
This shows that, indeed, the $SU(2)$ triplet of states rotates amongst itself.

\section{Appendix:  $M_{\zeta\eta}(x)$}\label{Mmn}

The integral of a meromorphic $(m,0)$ form $\omega$ on an oriented $m$ real dimensional submanifold $D$ of $\mathbb C^{m}$ is defined by the  homology class of $D$. The manifold we compute the homology of is that of $\mathbb C^{m}$ minus complex codimension 1 hyperplanes defined by the poles of  $\omega$. Thus with our choice of homology $C$ and $\tilde C$ we have a well-defined integral.

\para
For the practical evaluation of our integral we can evaluate each variable in turn choosing to sum either inside or outside the contour. This leads to many different ways of evaluating the integral, but the answer at the end of the calculation is always defined and independent of these choices as our homology class is well-defined. 

\para
For the evaluation of this integral we have been unable to prove that the residue has the value we claim it does for general $n,\zeta$ and $\eta$, but we have extensive numerical evidence for small values of $n$.

\para

Numerical evidence seems to imply that if we change the homology class to a homology class defined by the interleaving of $\zeta$ and $\eta$ then there is a unique pole contributing to the value of the integral given by the values
\be
w_l=q^{2s_l}/x\,,\quad\tilde w_l=q^{2\tilde s_l}/x\,.
\ee

\para

This homology class is defined as follows: the interleaving of $\zeta$ and $\eta$ allows us to define an ordering on the variables $W,\tilde W$ via $w_{i\alpha}\leqslant w_{j\beta}$ and $\tilde w_{i\alpha}\leqslant \tilde w_{j\beta}$ if $i<j$ or $i=j$ and $\alpha<\beta$ and $w_{i\alpha}\leqslant \tilde w_{j\beta}$ if $i\leqslant j$ and $\tilde w_{i\alpha}\leqslant w_{j\beta}$ if $i<j$. More than this we have the sets $P\subseteq\{w_1,\dots,w_n\}$ and $Q\subseteq\{\tilde w_1,\dots,\tilde w_n\}$ as the elements where the respective corresponding $s_l$ and $\tilde s_l$ are positive and their respective complements $\bar P$ and $\bar Q$ the elements where the respective corresponding $s_l$ and $\tilde s_l$ are negative. The contour is then defined by the elements of $P\cup Q$ size ordered by the ordering defined above with the smallest element having the smallest contour and the largest element having the largest contour, while for the elements of $\bar P\cup\bar Q$ we do the opposite ordering with the smallest element having the largest contour and so on. For evaluating the integral we sum all the poles of $P$ and $Q$ on the inside of the contour and outside the contour for $\bar P$ and $\bar Q$.

\para

With this choice of contour one finds that upon splitting the form we integrate into two parts, namely
\be\label{integrandpart}
M_{\zeta\,\eta}(x)=\frac1{N_\eta(q^2)(1-q^2)^n}\prod_lw_l^{\zeta_l}\tilde w_l^{-\eta_l}\times (\star)\,.
\ee
 The left over part represented by $\star$ contains all the relevant pole structure and evaluates to one on the pole, while the term to the left of $\star$ evaluates to exactly the answer required on the pole.

\para
Indeed on this pole we have that
\be
\prod_{l=1}^n w_l^{\zeta_l}\tilde w_l^{-\eta_l}=x^{|\eta|-|\zeta|}q^{\Delta_1[\zeta,\eta]}\,,\nonumber
\ee
where
\be\label{delta1}
\Delta_1[\zeta,\eta]:=\Delta_H[\zeta,\eta]-\frac{1}{2}\Delta_V[\eta]-\frac{1}{2}\Delta_V[\eta]\,,
\ee
with
\be
\Delta_H[\zeta,\eta]&:=&\sum_{l,m=1}^n|\zeta_l-\eta_m|\,,\nonumber\\
\Delta_V[\zeta]&:=&\sum_{l,m=1}^n|\zeta_l-\zeta_m|\,.\nonumber
\ee
This is because
\be
\Delta_1[\zeta,\eta]&=&\frac{1}{2}\sum_{l,m=1}^n\Big(2|\zeta_l-\eta_m|-|\zeta_l-\zeta_m|-|\eta_l-\eta_m|\Big)\nonumber\\
&=&\sum_{i\leqslant j}\sum_{\alpha_i=1}^{n_i}\sum_{\tilde\alpha_j=1}^{\tilde n_j}(\zeta_{i,\alpha_i}-\eta_{j,\tilde\alpha_j})+\sum_{j<i}\sum_{\alpha_i=1}^{n_i}\sum_{\tilde\alpha_j=1}^{\tilde n_j}(\eta_{j,\tilde\alpha_j}-\zeta_{i,\alpha_i})\nonumber\\
&&-\sum_{i<j}\sum_{\alpha_i=1}^{n_i}\sum_{\alpha_j=1}^{n_j}(\zeta_{i,\alpha_i}-\zeta_{j,\alpha_j})-\sum_{i=1}^{i_{\tiny{\mbox{max}}}}\sum_{\alpha_i<\alpha'_i}(\zeta_{i,\alpha_i}-\zeta_{i,\alpha'_i})\nonumber\\
&&-\sum_{i<j}\sum_{\tilde\alpha_i=1}^{\tilde n_i}\sum_{\tilde\alpha_j=1}^{\tilde n_j}(\eta_{i,\tilde\alpha_i}-\eta_{j,\tilde\alpha_j})-\sum_{i=1}^{i_{\tiny{\mbox{max}}}}\sum_{\tilde\alpha_i<\tilde\alpha'_i}(\eta_{i,\tilde\alpha_i}-\eta_{i,\tilde\alpha'_i})\nonumber\\
&=&\sum_{i=1}^{i_{\tiny{\mbox{max}}}}\sum_{\alpha_i=1}^{n_i}2s_{i,\alpha_i}\zeta_{i,\alpha_i}-\sum_{i=1}^{i_{\tiny{\mbox{max}}}}\sum_{\tilde\alpha_i=1}^{\tilde n_i}2\tilde s_{i,\tilde\alpha_i}\eta_{i,\tilde\alpha_i}\,.\nonumber
\ee
This is exactly the contribution from the pole to the exponent of $q$. 

\para
However for the original homology choice it is more complicated. For example if we take $n=2$ and $(\zeta;\eta)=(-1,-3;-2,-4)$ but with the original contour then there no longer is a non-zero residue at the point $(w_1,w_2;\tilde w_1,\tilde w_2)=(q/x,q/x,q/x,q/x)$ and instead one finds the value of the matrix element is given by the sum of the residue of the pole at $(q/x,q^3/x;q^3/x,q/x)$ and $(q/x,0;0,q/x)$. However the final value of $M_{\zeta\eta}(x)$ remains the same.

\para
A case where the two homology choices are the same is when $\zeta_l\geqslant \eta_m$ for all $l$ and $m$. We prove the value of $M_{\zeta\eta}(x)$ is indeed what we want in this simpler case.  $M_{\zeta\eta}(x)$ is
\be\nonumber
M_{\zeta\eta}(x)&=&\prod_{l=1}^n\oint_C\frac{dw_l}{2\pi i}\oint_{\tilde C}\frac{d\tilde w_l}{2\pi i \tilde w_l}\frac{w_l^{\zeta_l}}{w_l-q/x}\frac{\tilde w_l^{-\eta_l-1}}{1-qx\tilde w_l}\\
&&\prod_{l,m}\frac{\tilde w_m-q^2w_l}{\tilde w_m-w_l}\prod_{l<m}\left(\frac{1-\tilde w_m/\tilde w_l}{1-q^2\tilde w_m/\tilde w_l}\frac{1-w_l/w_m}{1-q^2w_l/w_m}\right)\,.\nonumber
\ee
\para

First we need to argue that taking the poles at zero for the $w_l$ will lead to zero residue and so we can ignore these poles. We see this by considering evaluating the poles of the $\tilde w_l$'s before the $w_l$'s. $\tilde w_n$ only has poles at $w_l$, which then leads to $\tilde w_{n-1}$ only having poles at $w_{l'}$ for $l'\neq l$. This gives $\tilde w_l=w_{\sigma(l)}$ for some $\sigma\in S_n$, undoing our splitting into two integrals and giving us our original integral as in equation (\ref{originalM}). The important point to note is that now each $w_l$ has lowest power in its Taylor expansion $\zeta_l-\eta_{\sigma^{-1}(l)}\geqslant0$ (the $-1$ exponent goes away because of the numerator $\tilde w_{\sigma^{-1}(l)}-q^2w_l\mapsto (1-q^2)w_l$) and so there can't be a pole with any coordinate at zero.
\para

With this in mind we can now proceed to evaluating our integral. For $w_1$ there is a pole at $0$ and a pole at $q/x$. Since we are ignoring the pole at $0$ we need only consider the pole at $q/x$, the residue of which is
\be
M_{\zeta\eta}(x)&=&q^{\zeta_1}x^{-\zeta_1}\prod_{l=2}^n\oint\frac{dw_l}{2\pi i}\frac{w_l^{\zeta_l}}{w_l-q^3/x}\prod_l\oint\frac{d\tilde w_l}{2\pi i \tilde w_l}\frac{\tilde w_l^{-\eta_l-1}}{1-qx\tilde w_l}\nonumber\\
&&\prod_{l=2}^n\prod_{m}\frac{\tilde w_m-q^2w_l}{\tilde w_m-w_l}\prod_l\frac{\tilde w_l-q^3/x}{\tilde w_l-q/x}\prod_{l<m}\frac{1-\tilde w_m/\tilde w_l}{1-q^2\tilde w_m/\tilde w_l}\prod_{l<m,2}^n\frac{1-w_l/w_m}{1-q^2w_l/w_m}\,.\nonumber
\ee
Note that this is an equality as we know the pole at zero will have no contribution. For $w_2$ there is a pole at $0$ and at $q^3/x$. The pole at $q^3/x$ has residue
\be
M_{\zeta\eta}(x)&=&q^{\zeta_1+3\zeta_2}x^{-\zeta_1-\zeta_2}\prod_{l=3}^n\oint\frac{dw_l}{2\pi i}\frac{w_l^{\zeta_l}}{w_l-q^5/x}\prod_l\oint\frac{d\tilde w_l}{2\pi i \tilde w_l}\frac{\tilde w_l^{-\eta_l-1}}{1-qx\tilde w_l}\nonumber\\
&&\prod_{l=3}^n\prod_{m}\frac{\tilde w_m-q^2w_l}{\tilde w_m-w_l}\prod_l\frac{\tilde w_l-q^5/x}{\tilde w_l-q/x}\prod_{l<m}\frac{1-\tilde w_m/\tilde w_l}{1-q^2\tilde w_m/\tilde w_l}\prod_{l<m,3}^n\frac{1-w_l/w_m}{1-q^2w_l/w_m}\,.\nonumber
\ee
This continues with $w_r$ having a pole at $q^{2r-1}/x$ and at $0$, with us choosing the non-zero pole giving residue
\be
M_{\zeta\eta}(x)&=&q^{\zeta_1+\dots+(2r-1)\zeta_r}x^{-\zeta_1-\dots-\zeta_r}\prod_{l=r+1}^n\oint\frac{dw_l}{2\pi i}\frac{w_l^{\zeta_l}}{w_l-q^{2r+1}/x}\prod_l\oint\frac{d\tilde w_l}{2\pi i \tilde w_l}\frac{\tilde w_l^{-\eta_l-1}}{1-qx\tilde w_l}\nonumber\\
&&\prod_{l=r+1}^n\prod_{m}\frac{\tilde w_m-q^2w_l}{\tilde w_m-w_l}\prod_l\frac{\tilde w_l-q^{2r+1}/x}{\tilde w_l-q/x}\prod_{l<m}\frac{1-\tilde w_m/\tilde w_l}{1-q^2\tilde w_m/\tilde w_l}\prod_{l<m,r+1}^n\frac{1-w_l/w_m}{1-q^2w_l/w_m}\,.\nonumber
\ee
Finally we evaluate $w_n$ at $q^{2n-1}/x$, ignoring the pole at $0$ to get
\be
M_{\zeta\eta}(x)&=&q^{\zeta_1+\dots+(2n-1)\zeta_{n}}x^{-\zeta_1-\dots-\zeta_n}\prod_l\oint\frac{d\tilde w_l}{2\pi i \tilde w_l}\frac{\tilde w_l^{-\eta_l-1}}{1-qx\tilde w_l}\prod_l\frac{\tilde w_l-q^{2n+1}/x}{\tilde w_l-q/x}\prod_{l<m}\frac{1-\tilde w_m/\tilde w_l}{1-q^2\tilde w_m/\tilde w_l}\,.\nonumber
\ee
The only pole for $\tilde w_n$ is then $q/x$. Ignoring the factors of $q$ and $x$ at the front from the previous line this gives
\be
M_{\zeta\eta}(x)&\propto&q^{-\eta_n-1}x^{\eta_n+1}\prod_{l=1}^{n-1}\oint\frac{d\tilde w_l}{2\pi i \tilde w_l}\frac{\tilde w_l^{-\eta_l-1}}{1-qx\tilde w_l}\frac{1}{1-q^2}\nonumber\\
&&\prod_{l=1}^{n-1}\frac{\tilde w_l-q^{2n+1}/x}{\tilde w_l-q^3/x}(q/x-q^{2n+1}/x)\prod_{l<m,1}^{n-1}\frac{1-\tilde w_m/\tilde w_l}{1-q^2\tilde w_m/\tilde w_l}\nonumber
\ee
where $\propto$ is used instead of an equals sign because we are ignoring the powers of $q$ and $x$ arising from the $w$ integrals. The only pole for $\tilde w_{n-1}$ is $q^3/x$ with residue
\be
M_{\zeta\eta}(x)&\propto&q^{-\eta_n-3\eta_{n-1}-3}x^{\eta_n+\eta_{n-1}+1}\prod_{l=1}^{n-2}\oint\frac{d\tilde w_l}{2\pi i \tilde w_l}\frac{\tilde w_l^{-\eta_l-1}}{1-qx\tilde w_l}\frac{1}{1-q^2}\frac1{1-q^4}\prod_{l=1}^{n-2}\frac{\tilde w_l-q^{2n+1}/x}{\tilde w_l-q^5/x}\nonumber\\
&&(q^3/x-q^{2n+1}/x)(1-q^{2n})\prod_{l<m,1}^{n-2}\frac{1-\tilde w_m/\tilde w_l}{1-q^2\tilde w_m/\tilde w_l}\,.\nonumber
\ee
This continues with the only contributing pole for $\tilde w_{n-r+1}$ being $q^{2r-1}/x$ with residue
\be
M_{\zeta\eta}(x)&\propto&q^{-\sum_{l=n-r+1}^n(2n-2l+1)\eta_l}x^{\sum_{l=n-r+1}^n\eta_l}\prod_{l=1}^{n-r}\oint\frac{d\tilde w_l}{2\pi i \tilde w_l}\frac{\tilde w_l^{-\eta_l-1}}{1-qx\tilde w_l}\prod_{l=1}^r\frac{1}{1-q^{2l}}\nonumber\\
&&\prod_{l=1}^{n-r}\frac{\tilde w_l-q^{2n+1}/x}{\tilde w_l-q^{2r+1}/x}\prod_{l=n-r+1}^n(1-q^{2l})\prod_{l<m,1}^{n-r}\frac{1-\tilde w_m/\tilde w_l}{1-q^2\tilde w_m/\tilde w_l}\,.\nonumber
\ee
Taking $\tilde w_1=q^{2n-1}/x$ gives our expected result.

\section{Appendix: Evaluating $\mathcal O_{\zeta\eta}(k)$}\label{Omn}

We note that $\mathcal O_{\zeta\eta}(k)=\mathcal O_{\zeta-(k^n),\eta}(0)$, so it is sufficient to evaluate $\mathcal O_{\zeta\eta}(0)\equiv\mathcal O_{\zeta\eta}$. Using the permutation invariance of the integrand we can rewrite $\mathcal O_{\zeta\eta}$ as
\be
\mathcal O_{\zeta\eta}&=&\frac{(1-q^2)^n}{N_\eta(q^2)}\prod_{l=1}^n\left(\oint_C\frac{dw_l}{2\pi i}\oint_C\frac{d\tilde w_l}{2\pi i}\right)\prod_{l=1}^nw_l^{\zeta_l}\tilde w_l^{-\eta_l}\prod_{l<m}(w_m-w_l)(\tilde w_m-q^2\tilde w_l)\nonumber\\
&&\prod_{l>m}(\tilde w_m-\tilde w_l)(w_m-q^2w_l)\prod_{l,m}\frac{1}{(w_m-qz\tilde w_l)(\tilde w_m-\frac qzw_l)}\,.\nonumber
\ee
We define $n_+,\tilde n_+$ such that $\eta_{n_+},\zeta_{\tilde n_+}\geqslant 0$ and $\eta_{n_++1},\zeta_{\tilde n_++1}<0$. Note that $w_1,\dots,w_{n_+}$ only have poles within the unit circle at $qz\tilde w_l$ while $\tilde w_{\tilde n_++1},\dots,\tilde w_n$ only have poles within the unit circle at $\frac qzw_l$. 
We do these integrals defining $\sigma:\{1,\dots,n_+\}\to\{1,\dots,n\}$ and $\tau:\{\tilde n_++1,\dots,n\}\to\{1,\dots,n\}$
so $w_l=wz\tilde w_{\sigma(l)}$  for $l=1,\dots,n_+$ and $\tilde w_l=\frac qzw_{\tau(l)}$ for $l=\tilde n_++1,\dots,n$.
Note that the factor $\prod_{l<m}(w_m-w_l)(\tilde w_l-\tilde w_m)$ guarantees that $\sigma$ and $\tau$ are injective.

\para
Now we show that
\be
\sigma\{1,\dots,n_+\}&\subseteq&\{1,\dots,\tilde n_+\}\,,\nonumber\\
\tau\{\tilde n_++1,\dots,n\}&\subseteq&\{n_++1,\dots,n\}\,,\nonumber
\ee
otherwise the integral is zero. We prove this for $\sigma$ and note that a very similar proof will work for $\tau$. Suppose $\exists \,l\in\{1,\dots,n_+\}$ such that $\sigma(l)>\tilde n_+$ then $w_l=qz\tilde w_{\sigma(l)}=q^2w_{\tau\sigma(l)}$. Either $\tau\sigma(l)>l$ so the factor in the integrand $(1-q^2w_{\tau\sigma(l)}/w_l)$ evaluates to zero or $\tau\sigma(l)<l\leqslant n_+\implies w_{\tau\sigma(l)}=qz\tilde w_{\sigma\tau\sigma(l)}\implies\tilde w_{\sigma(l)}=q^2\tilde w_{\sigma\tau\sigma(l)}$, now either $\sigma(l)>\sigma\tau\sigma(l)$ so the factor $(\tilde w_{\sigma(l)}-q^2\tilde w_{\sigma\tau\sigma(l)})$ evaluates to zero or $\tilde w_{\sigma\tau\sigma(l)}=\frac qzw_{\tau\sigma\tau\sigma(l)}$. This process carries on, either we hit a zero, or we hit a fixed point of $\tau\sigma$ (as it is a permutation of a finite set), however this would lead to $w_m=q^2w_m$ for some $m$ with $w_m\neq 0$, a clear contradiction.

Now we show that if $\tilde n_+>n_+$ then we get zero. Suppose that this is true, we extend our definition of $\tau$ to $\{n_++1,\dots,n\}$ in order that $\tau\{n_++1,\dots,n\}=\{n_++1,\dots,n\}$ and then extend our definition of $\sigma$ such that $\{\sigma(1),\dots,\sigma(n_+),\sigma\tau(n_++1),\dots,\sigma\tau(\tilde n_+)\}=\{1,\dots,\tilde n_+\}$. After doing these integrals  we consider $w_{\tau(n_++1)},\dots,w_{\tau(\tilde n_+)}$. As $m_{\tau(l)}<0$ we consider the poles outside the unit circle, there is only one and it is at $w_{\tau(l)}=\frac zq\tilde w_m$ for $m=1,\dots,\tilde n_+$. The factor $\prod_{l=1}^{n_+}\prod_{n_++1}^n(\tilde w_{\sigma(l)}-\frac qzw_m)$ guarantees that the pole is $w_{\tau(l)}=\frac zq\tilde w_{\sigma\tau(l)}$ for $l=n_++1,\dots,\tilde n_+$.

Upon doing this integral one finds the integrand
\be
\mathcal O_{\zeta\eta}&=&\frac{(-)^{\#}z^{\#}q^{\#}}{N_\eta(q^2)}\prod_{l=\tilde n_++1}^n\oint\frac{dw_{\tau(l)}}{2\pi iw_{\tau(l)}}\prod_{l=1}^{\tilde n_+}\oint\frac{d\tilde w_l}{2\pi i\tilde w_l}\nonumber\\
&&\prod_{l=\tilde n_++1}^nw_{\tau(l)}^{\zeta_{\tau(l)}-\eta_l}\prod_{l=1}^{n_+}\tilde w_{\sigma(l)}^{\zeta_l-\eta_{\sigma(l)}}\prod_{l=n_++1}^{\tilde n_+}\tilde w_{\sigma\tau(l)}^{\zeta_{\tau(l)}-\eta_{\sigma\tau(l)}}\nonumber
\ee

This means that $\tilde w_1,\dots, \tilde w_{\tilde n_+}$ can only receive poles within the unit circle from the monomial at zero, and will have a non-zero residue only if it is a simple pole. However we have that $\zeta_{\tau(l)}-\eta_{\sigma\tau(l)}-1\leqslant -2$ for $l=n_++1,\dots,\tilde n_+$ because $\zeta_{\tau(l)}<0$ and $\eta_{\sigma\tau(l)}\geqslant0$ for such $l$.
 Hence this cannot be a simple pole. Thus we may conclude that the integral will evaluate to zero unless $\tilde n_+\leqslant n_+$. We now do the integral for $w_1,\dots,w_{n_+}=qzw_{\sigma(1)},\dots,qzw_{\sigma(n_+)}$ and $\tilde w_{n_++1},\dots,\tilde w_n=\frac qzw_{\tau(n_++1)},\dots,\frac qzw_{\tau(n)}$. By our earlier argument we have that $\sigma\{1,\dots,n_+\}=\{1,\dots,n_+\}$ and $\tau\{n_++1,\dots,n\}=\{n_++1,\dots,n\}$, so we may combine $\sigma$ and $\tau$ together to form $\sigma\in S_{n_+}\times S_{n-n_+}\subseteq S_n$. Upon performing the integral we obtain that all the cross terms cancel out. We define the dot product of the Hall-Littlewood polynomials for $r\leqslant n$ variables to be
\be
\langle\Psi_\theta,\Psi_{\theta'}\rangle^{(r)}&:=&\frac{1}{N_\theta N_{\theta'}}\prod_{i=1}^r\oint\frac{dw_i}{2\pi iw_i}\sum_{\sigma\in S_r}\prod_{i=1}^rw_i^{\theta_i-\theta'_{\sigma^{-1}(i)}}\prod_{i<j}(1-q^2w_j/w_i)(1-w_{\sigma(j)}/w_{\sigma(i)})\nonumber\\
&&\prod_{i>j}(1-q^2w_{\sigma(j)}/w_{\sigma(i)})(1-w_j/w_i)\prod_{i\neq j}\frac{1}{(1-q^2w_j/w_i)(1-w_j/w_i)}\nonumber\\
&=&\delta_{\theta\theta'}\frac{1}{N^{(r)}_\theta(q^2)}\,,\nonumber
\ee
where $N^{(r)}$ has an $r$ to indicate it is for partitions of size $r$ and not $n$. We find that the integral becomes, defining $\zeta_{(n_+)}:=(\zeta_1,\dots,\zeta_{n_+})$ and $\zeta^{(n_+)}:=(\zeta_{n_++1},\dots,\zeta_n)$ (and similarly for $\eta$),
\be
\mathcal O_{\zeta\eta}&=&N_\zeta q^{\sum_1^{n_+}\zeta_l-\sum_{n_++1}^n\eta_l}z^{\sum_1^{n_+}\zeta_l+\sum_{n_++1}^n\eta_l}\langle\Psi_{\zeta_{(n_+)}},\Psi_{\eta_{(n_+)}}\rangle^{(n_+)}\langle\Psi_{\zeta^{(n_+)}},\Psi_{\eta^{(n_+)}}\rangle^{(n-n_+)}\nonumber\\
&=&\delta_{\zeta\eta}z^{|\eta|}q^{||\eta||}\,\nonumber
\ee
as required.

\section*{Acknowledgements}

We are supported  by the European Research Council under the European
Union's Seventh Framework Programme (FP7/2007-2013), ERC grant
agreement STG 279943, ``Strongly Coupled Systems", and by  the STFC consolidated grant ST/P000681/1.


\begin{thebibliography}{99}

\small
\parskip=0pt plus 2pt

\bibitem{zhang} S.~C.~Zhang and J.~P.~Hu,
  ``{\it A Four-dimensional generalization of the quantum Hall effect},''
  Science {\bf 294}, 823 (2001)
  [cond-mat/0110572].
  
  
\bibitem{zhang2}   J.~P.~Hu and S.~C.~Zhang,
  ``{\it Collective excitations at the boundary of a 4-D quantum Hall droplet},''
  Phys.\ Rev.\ B {\bf 66}, 125301 (2002)
  [cond-mat/0112432].

\bibitem{henriette}   H.~Elvang and J.~Polchinski, ``{\it The Quantum Hall effect on $R^4$}''
  hep-th/0209104.
  
  
  \bibitem{fabinger}   M.~Fabinger,
  ``{\it Higher dimensional quantum Hall effect in string theory}''
  JHEP {\bf 0205}, 037 (2002)
  [hep-th/0201016].
  
  \bibitem{bernevig}   B.~A.~Bernevig, C.~H.~Chern, J.~P.~Hu, N.~Toumbas and S.~C.~Zhang,
  ``{\it Effective field theory description of the higher dimensional quantum Hall liquid},''
  Annals Phys.\  {\bf 300}, 185 (2002)
  [cond-mat/0206164].
  
  \bibitem{chen} Y.~X.~Chen, B.~Y.~Hou and B.~Y.~Hou,
  ``{\it Noncommutative geometry of four-dimensional quantum Hall droplet},''
  Nucl.\ Phys.\ B {\bf 638}, 220 (2002)
  [hep-th/0203095].
  
  
 \bibitem{knair1}  D.~Karabali and V.~P.~Nair,
  ``{\it Quantum Hall effect in higher dimensions},''
  Nucl.\ Phys.\ B {\bf 641}, 533 (2002)
  [hep-th/0203264].
 
 \bibitem{knair2}   D.~Karabali and V.~P.~Nair,
  ``{\it Edge states for quantum Hall droplets in higher dimensions and a generalized WZW model},''
  Nucl.\ Phys.\ B {\bf 697}, 513 (2004)
  [hep-th/0403111].
  
  
 \bibitem{knair3}   D.~Karabali and V.~P.~Nair,
  ``{\it Quantum Hall effect in higher dimensions, matrix models and fuzzy geometry},''
  J.\ Phys.\ A {\bf 39}, 12735 (2006)
  [hep-th/0606161].
  
  \bibitem{knair4}   D.~Karabali and V.~P.~Nair,
  ``{\it Geometry of the quantum Hall effect: An effective action for all dimensions},''
  Phys.\ Rev.\ D {\bf 94} (2016) no.2,  024022
  [arXiv:1604.00722 [hep-th]].
  
\bibitem{eight}   B.~A.~Bernevig, J.~p.~Hu, N.~Toumbas and S.~C.~Zhang,
  ``{\it The Eight-dimensional quantum Hall effect and the octonions},''
  Phys.\ Rev.\ Lett.\  {\bf 91}, 236803 (2003)
  [cond-mat/0306045].
  
 \bibitem{heckman}  J.~J.~Heckman and L.~Tizzano,
  ``6D Fractional Quantum Hall Effect,''
  arXiv:1708.02250 [hep-th].
  

\bibitem{cold1} H. M. Price, O.  Zilberberg, T. Ozawa, I.  Carusotto, N.  Goldman, 
  ``{\it  Four-Dimensional Quantum Hall Effect with Ultracold Atoms}",  Phys. Rev. Lett. 115, 195303 (2015), 
  arXiv:1505.04387 [cond-mat.quant-gas]
  
 \bibitem{cold2} M. Lohse, C. Schweizer, H. M. Price, O, Zilberberg, I. Bloch, ``{\it Exploring 4D Quantum Hall Physics with a 2D Topological Charge Pump}", 	arXiv:1705.08371 [cond-mat.quant-gas]
  
  
 \bibitem{yang}   C.~N.~Yang,
  ``{\it Generalization of Dirac's Monopole to SU(2) Gauge Fields},''
  J.\ Math.\ Phys.\  {\bf 19}, 320 (1978).
  
  
  

\bibitem{bw}  B.~Blok and X.~G.~Wen,
  ``{\it Many body systems with non-Abelian statistics},''
  Nucl.\ Phys.\ B {\bf 374}, 615 (1992).
  

  

\bibitem{nek}   N.~A.~Nekrasov,
  ``{\it Seiberg-Witten prepotential from instanton counting},''
  Adv.\ Theor.\ Math.\ Phys.\  {\bf 7}, no. 5, 831 (2003)
  [hep-th/0206161].

\bibitem{nekok1} N.~Nekrasov and A.~Okounkov,
  ``{\it Seiberg-Witten theory and random partitions},''
  Prog.\ Math.\  {\bf 244}, 525 (2006)
  [hep-th/0306238].


\bibitem{yuji}  Y.~Tachikawa,
  ``{\it Five-dimensional Chern-Simons terms and Nekrasov's instanton counting},''
  JHEP {\bf 0402}, 050 (2004)
  [hep-th/0401184].



\bibitem{ami1}  S.~Cremonesi, A.~Hanany and A.~Zaffaroni,
  ``{\it Monopole operators and Hilbert series of Coulomb branches of $3d$  $\mathcal{N} = 4$ gauge theories},''
  JHEP {\bf 1401}, 005 (2014)
  [arXiv:1309.2657 [hep-th]].

 
  \bibitem{ami2}  S.~Cremonesi, A.~Hanany, N.~Mekareeya and A.~Zaffaroni,
  ``{\it Coulomb branch Hilbert series and Hall-Littlewood polynomials},''
  JHEP {\bf 1409}, 178 (2014)
  [arXiv:1403.0585 [hep-th]].


\bibitem{ami3}   S.~Cremonesi, G.~Ferlito, A.~Hanany and N.~Mekareeya,
  ``{\it Coulomb Branch and The Moduli Space of Instantons},''
  JHEP {\bf 1412}, 103 (2014)
  [arXiv:1408.6835 [hep-th]].
  




 \bibitem{alexios}
A.~P.~Polychronakos,
``{\it Quantum Hall states as matrix Chern-Simons theory}''
JHEP {\bf 0104}, 011 (2001)
[arXiv:hep-th/0103013].

\bibitem{us2}
N.~Dorey, D.~Tong and C.~Turner,
  ``{\it Matrix model for non-Abelian quantum Hall states},''
  Phys.\ Rev.\ B {\bf 94}, no. 8, 085114 (2016)
  [arXiv:1603.09688 [cond-mat.str-el]].

\bibitem{susskind}   L.~Susskind,
  ``{\it The Quantum Hall fluid and noncommutative Chern-Simons theory},''
  hep-th/0101029.

\bibitem{unknown}   D.~Tong,
  ``{\it A Quantum Hall fluid of vortices},''
  JHEP {\bf 0402}, 046 (2004)
  [hep-th/0306266].

  \bibitem{us1} D.~Tong and C.~Turner,
  ``{\it Quantum Hall effect in supersymmetric Chern-Simons theories},''
  Phys.\ Rev.\ B {\bf 92}, no. 23, 235125 (2015)
  [arXiv:1508.00580 [hep-th]].





\bibitem{hellvram}
S.~Hellerman and M.~Van Raamsdonk,
``{\em Quantum Hall physics equals noncommutative field theory}''
JHEP {\bf 0110}, 039 (2001)
[arXiv:hep-th/0103179].

\bibitem{ks}
D.~Karabali and B.~Sakita, 
``{\em Chern-Simons matrix model: Coherent states and relation to Laughlin  wavefunctions}'', 
Phys.\ Rev.\ B {\bf 64}, 245316 (2001)
[arXiv:hep-th/0106016]. \\ 

\bibitem{ks2}D.~Karabali and B.~Sakita,
``{\em Orthogonal basis for the energy eigenfunctions of the Chern-Simons  matrix model}''
Phys.\ Rev.\ B {\bf 65}, 075304 (2002)
[arXiv:hep-th/0107168].


\bibitem{hkk} T. H. Hansson, J. Kailasvuori and A. Karlhede, 
``{\em Charge and current in the quantum Hall matrix model}'', arXiv:cond-mat/0304271.

\bibitem{cappelli}  A.~Cappelli and I.~D.~Rodriguez,
  JHEP {\bf 0612}, 056 (2006)
  [hep-th/0610269].

\bibitem{cappelli2}   A.~Cappelli and M.~Riccardi,
  ``{\it Matrix model description of Laughlin Hall states},''
  J.\ Stat.\ Mech.\  {\bf 0505}, P05001 (2005)
  [hep-th/0410151].

\bibitem{edge}   I.~D.~Rodriguez,
  ``{\it Edge excitations of the Chern Simons matrix theory for the FQHE},''
  JHEP {\bf 0907}, 100 (2009)
  [arXiv:0812.4531 [hep-th]].


\bibitem{us3}   N.~Dorey, D.~Tong and C.~Turner,
  ``{\it A Matrix Model for WZW},''
  JHEP {\bf 1608}, 007 (2016)
  [arXiv:1604.05711 [hep-th]].








\bibitem{seib}   N.~Seiberg,
  ``{\it Five-dimensional SUSY field theories, nontrivial fixed points and string dynamics},''
  Phys.\ Lett.\ B {\bf 388}, 753 (1996)
  [hep-th/9608111].

 
 \bibitem{kimyeong1} S.~Kim and S.~Lee,
  ``{\it The Geometry of dyonic instantons in 5-dimensional supergravity},''
  arXiv:0712.0090 [hep-th].
 
 \bibitem{collie}  B.~Collie and D.~Tong,
  ``{\it Instantons, Fermions and Chern-Simons Terms},''
  JHEP {\bf 0807}, 015 (2008)
  [arXiv:0804.1772 [hep-th]].
 
 \bibitem{kimyeong2}  S.~Kim, K.~M.~Lee and S.~Lee, ``{\it Dyonic Instantons in 5-dim Yang-Mills Chern-Simons Theories},''
  JHEP {\bf 0808}, 064 (2008)
  [arXiv:0804.1207 [hep-th]].
 
 
 
\bibitem{nairs1}   V.~P.~Nair and J.~Schiff,
  ``{\it A Kahler-{Chern-Simons} Theory and Quantization of Instanton Moduli Spaces},''
  Phys.\ Lett.\ B {\bf 246}, 423 (1990).
 
 \bibitem{nairs2}   V.~P.~Nair and J.~Schiff,
  ``{\it Kahler Chern-Simons theory and symmetries of antiselfdual gauge fields},''
  Nucl.\ Phys.\ B {\bf 371}, 329 (1992).
 
 
 \bibitem{adhm}   M.~F.~Atiyah, N.~J.~Hitchin, V.~G.~Drinfeld and Y.~I.~Manin,
  ``{\it Construction of Instantons},''
  Phys.\ Lett.\ A {\bf 65}, 185 (1978).
 
  
 \bibitem{douglas}  M.~R.~Douglas,
  ``{\it Gauge fields and D-branes},''
  J.\ Geom.\ Phys.\  {\bf 28}, 255 (1998)
  [hep-th/9604198].
  
  


 
 \bibitem{andrew} N.~Dorey and A.~Singleton,
  ``{\it Instantons, Integrability and Discrete Light-Cone Quantisation},''
  arXiv:1412.5178 [hep-th].
  
   \bibitem{nekschwarz}   N.~Nekrasov and A.~S.~Schwarz,
  ``{\it Instantons on noncommutative R**4 and (2,0) superconformal six-dimensional theory},''
  Commun.\ Math.\ Phys.\  {\bf 198}, 689 (1998)
  [hep-th/9802068].
  
 \bibitem{wrong1}   Y.~X.~Chen,
  ``{\it Matrix models of four-dimensional quantum Hall fluids},''
  hep-th/0209182.


\bibitem{wrong2}   Y.~X.~Chen,
  ``{\it Quasiparticle excitations and hierarchies of four-dimensional quantum Hall fluid states in the matrix models},''
  hep-th/0210059.
  
  \bibitem{liwu} Yi. Lu and C. Wu, ``{\it High-Dimensional Topological Insulators with Quaternionic Analytic Landau Levels}", 
  Phys. Rev. Lett. {\bf 110}, 216802 (2013), 	arXiv:1103.5422 [cond-mat.str-el]
  
  \bibitem{nakajima} H. Nakajima, ``{\it Lectures on Hilbert Scheme of Points on Surfaces}", Am. Math. Soc., (1999). 

\bibitem{ami67}  S.~Benvenuti, A.~Hanany and N.~Mekareeya,
  ``{\it The Hilbert Series of the One Instanton Moduli Space},''
  JHEP {\bf 1006}, 100 (2010)
  [arXiv:1005.3026 [hep-th]].

 

 
 
 \bibitem{nekok}  E.~Carlsson, N.~Nekrasov and A.~Okounkov,
  ``{\it Five dimensional gauge theories and vertex operators},''
  Moscow Math.\ J.\  {\bf 14}, no. 1, 39 (2014)
  [arXiv:1308.2465 [math.RT]].
  
  \bibitem{koreans} 
    H.~C.~Kim, S.~Kim, E.~Koh, K.~Lee and S.~Lee,
  ``{\it On instantons as Kaluza-Klein modes of M5-branes},''
  JHEP {\bf 1112}, 031 (2011)
  [arXiv:1110.2175 [hep-th]].

 
    \bibitem{intseib}   K.~A.~Intriligator and N.~Seiberg,
  ``{\it Mirror symmetry in three-dimensional gauge theories},''
  Phys.\ Lett.\ B {\bf 387}, 513 (1996)
  [hep-th/9607207].
  
  \bibitem{mirror}  J.~de Boer, K.~Hori, H.~Ooguri and Y.~Oz,
  ``{\it Mirror symmetry in three-dimensional gauge theories, quivers and D-branes},''
  Nucl.\ Phys.\ B {\bf 493}, 101 (1997)
  [hep-th/9611063].

\bibitem{Macdonald:95}
  I.G. Macdonald, Symmetric Functions and Hall Polynomials (2nd ed.), Oxford Uni- versity Press 1995; Lect. Note in Math. 1271 (1987) 189-200, Springer; Publ. I.R.M.A. (1988) 131-171.
  



\end{thebibliography}
\end{document}